\documentclass[letter]{aa} 
\usepackage{graphicx}
\usepackage{txfonts}
\usepackage{natbib}
\usepackage{amsmath}
\usepackage{color}
\usepackage{multicol}
\usepackage{longtable}
\usepackage{lscape}
\usepackage{enumerate}
\usepackage{mathtools}
\usepackage{lscape}
\begin{document} 

%-----------------------------------------------------------------------
% TITLE & AUTHORS
%-----------------------------------------------------------------------
   \title{The complexity of Orion: an ALMA view\thanks{This paper makes use of the following ALMA data: ADS/JAO.ALMA$\#$2013.1.00533.S . ALMA is a partnership of ESO (representing its member states), NSF (USA) and NINS (Japan), together with NRC (Canada), NSC and ASIAA (Taiwan), and KASI (Republic of Korea), in cooperation with the Republic of Chile. The Joint ALMA Observatory is operated by ESO, AUI/NRAO and NAOJ.}}

   \subtitle{II. gGg$^{\prime}$-Ethylene Glycol and Acetic Acid}
   
   \author{C. Favre
          \inst{1}
           \and
          L. Pagani\inst{2}          
          \and
          P. Goldsmith\inst{3}
          \and
          E. Bergin\inst{4}          
        \and 
          M. Carvajal\inst{5,6}
        \and 
          I. Kleiner\inst{7}
	 \and  
         G. Melnick\inst{8}
         \and  
         R. Snell\inst{9} 
	             }

   \institute{INAF--Osservatorio Astrofisico di Arcetri, Largo E. Fermi 5, Firenze, 50125, Italy\\
              \email{cfavre@arcetri.astro.it}
         \and
                LERMA, UMR 8112 du CNRS, Observatoire de Paris, 61 Av. de l'Observatoire, 75014 Paris, France
          %    \email{laurent.pagani@obspm.fr}
         \and
	      Jet Propulsion Laboratory, California Institute of Technology, 4800 Oak Grove Drive, Pasadena, CA 91109, USA
         \and
	      Department of Astronomy, University of Michigan, 500 Church Street, Ann Arbor, MI 48109, USA
         \and
             Dpto. Ciencias Integradas, Unidad GIFMAN-UHU Asociada al CSIC, Universidad de Huelva, 21071, Huelva, Spain
         \and
             Instituto Universitario Carlos I de F\'{\i}sica Te\'orica y Computacional, Universidad de Granada, Granada, Spain
  	 \and
          Laboratoire Interuniversitaire des Syst\`emes Atmosph\'eriques (LISA), CNRS, UMR 7583, Universit\'e de Paris-Est et Paris Diderot, 61, Av. du G\'en\'eral de Gaulle, F-94010 Cr\'eteil Cedex, France
         \and
             Harvard-Smithsonian Center for Astrophysics, Cambridge, Massachusetts, USA
         \and
             Department of Astronomy, University of Massachusetts, Amherst, MA, 01003, USA
            }

   \date{Received September 15, 1996; accepted March 16, 1997}

% \abstract{}{}{}{}{} 
% 5 {} token are mandatory
 
%===============================================================
%
%-----------------------------------------------------------------------------------------------------------------------------
%--------------------------------------------ABSTRACT -------------------------------------------------
%-----------------------------------------------------------------------------------------------------------------------------
%
  \abstract 
   {We report the first detection and high angular resolution (1.8$\arcsec$ $\times$ 1.1$\arcsec$) imaging of acetic acid (CH$_3$COOH) and gGg$^{\prime}$--ethylene glycol (gGg$^{\prime}$(CH$_2$OH)$_2$) towards the Orion Kleinmann--Low nebula. The observations were carried out at $\sim$1.3mm with ALMA during the Cycle~2. A notable result is that the spatial distribution of the acetic acid and ethylene glycol emission differs from that of the other O-bearing molecules within Orion-KL. Indeed, while the typical emission of O-bearing species harbors a morphology associated with a "V-shape" linking the Hot Core region to the Compact Ridge (with an extension towards the BN object), that of acetic acid and ethylene glycol mainly peaks at about 2$^{\arcsec}$ southwest from the hot core region (near sources I and n). We find that the measured CH$_3$COOH:aGg$^{\prime}$(CH$_2$OH)$_2$ and CH$_3$COOH:gGg$^{\prime}$(CH$_2$OH)$_2$ ratios differ from the ones measured towards the low-mass protostar IRAS 16293--2422 by 
more than one order of magnitude. Our best hypothesis to explain these findings is that CH$_3$COOH, aGg$^{\prime}$(CH$_2$OH)$_2$ and gGg$^{\prime}$(CH$_2$OH)$_2$ are formed on the icy-surface of grains and then released into the gas-phase, via co-desorption with water, due to a bullet of matter ejected during the explosive event that occurred in the heart of the Nebula about 500-700 years ago.  
 }
%  
  % ----------------------------------
% --- Keywords  6 max---
% ----------------------------------

   \maketitle
%
%===============================================================
%===============================================================
%
%-----------------------------------------------------------------------------------------------------------------------------
%--------INTRODUCTION --------------------------------------------
%------------------------------------------------------------------
\section{Introduction}
\label{sec:introduction}
%------------------------------------------------------------------
The Orion Kleinmann--Low nebula (hereafter Orion--KL) is the closest high mass-star forming region \citep[388$\pm$5~pc,][]{Kounkel:2017}. 
Its proximity and rich molecular composition make this region well-suited for astrochemical study. In this context, numerous single-dish surveys, including the broadband Herschel/HIFI HEXOS survey \citep[][]{Bergin:2010,Crockett:2014}, as well as interferometric observations have been performed towards this region \citep[e.g.][and references therein]{Favre:2015b,Pagani:2017}. It is important to note that two main molecular components are associated with Orion-KL: the Compact Ridge and the Hot Core. The latter region may have resulted from interaction of the surrounding gas with remnants of the explosive event -- triggered by the close encounter  of the sources I, n and the BN object --  that occurred in the region about 500--700 years ago \citep[e.g. see,][and references therein]{Zapata:2011,Nissen:2012}. 
Thus, the complex physical structure and history make Orion-KL an interesting source -- that may not be representative of other high-mass star forming regions -- to study the production route (at the icy surface of grains and/or in the gas phase) of complex organic molecules \citep[i.e. molecules that contain six or more atomes, including carbon, hereafter COMs, see][]{Herbst:2009}. 
Although present in other star-forming regions, some COMs have not yet been detected in Orion-KL
This mainly results from sensitivity limitation and a high spectral confusion level \citep[e.g. see,][]{Tercero:2010}.
High resolution and sensitivity, as offered by ALMA, are thus mandatory to search for weak lines associated with COMs. In that context, we have used ALMA during Cycle~2 to perform deep observations of this region in a fraction of band 6 ($\approx$ 1-2 mm).

%------------------------------------------------------------------
% --- FIGURE 1 ---
%-----------------------------------------------------------------
\begin{figure*}[h!]
\centering
\includegraphics[width=6cm]{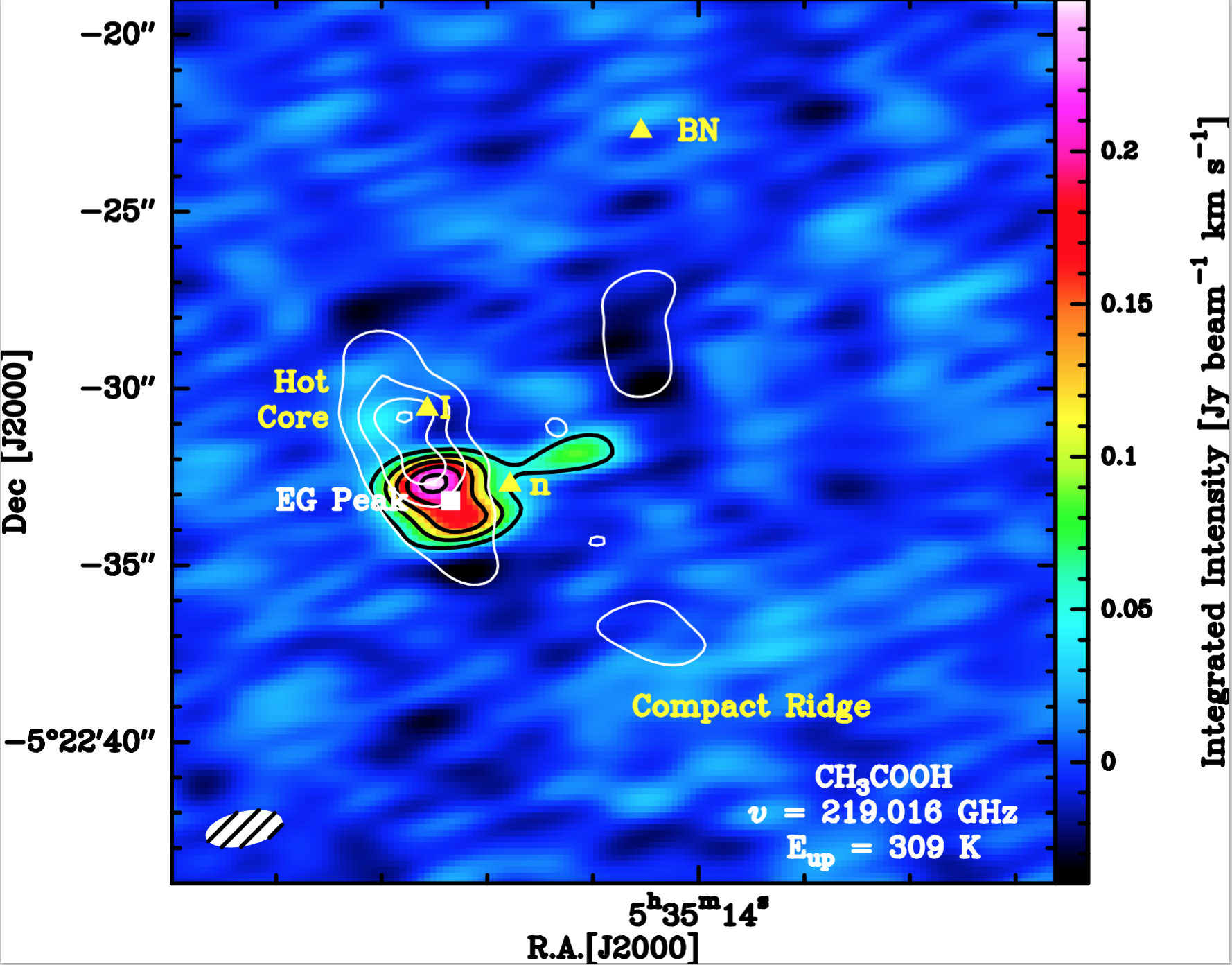}
\includegraphics[width=6cm]{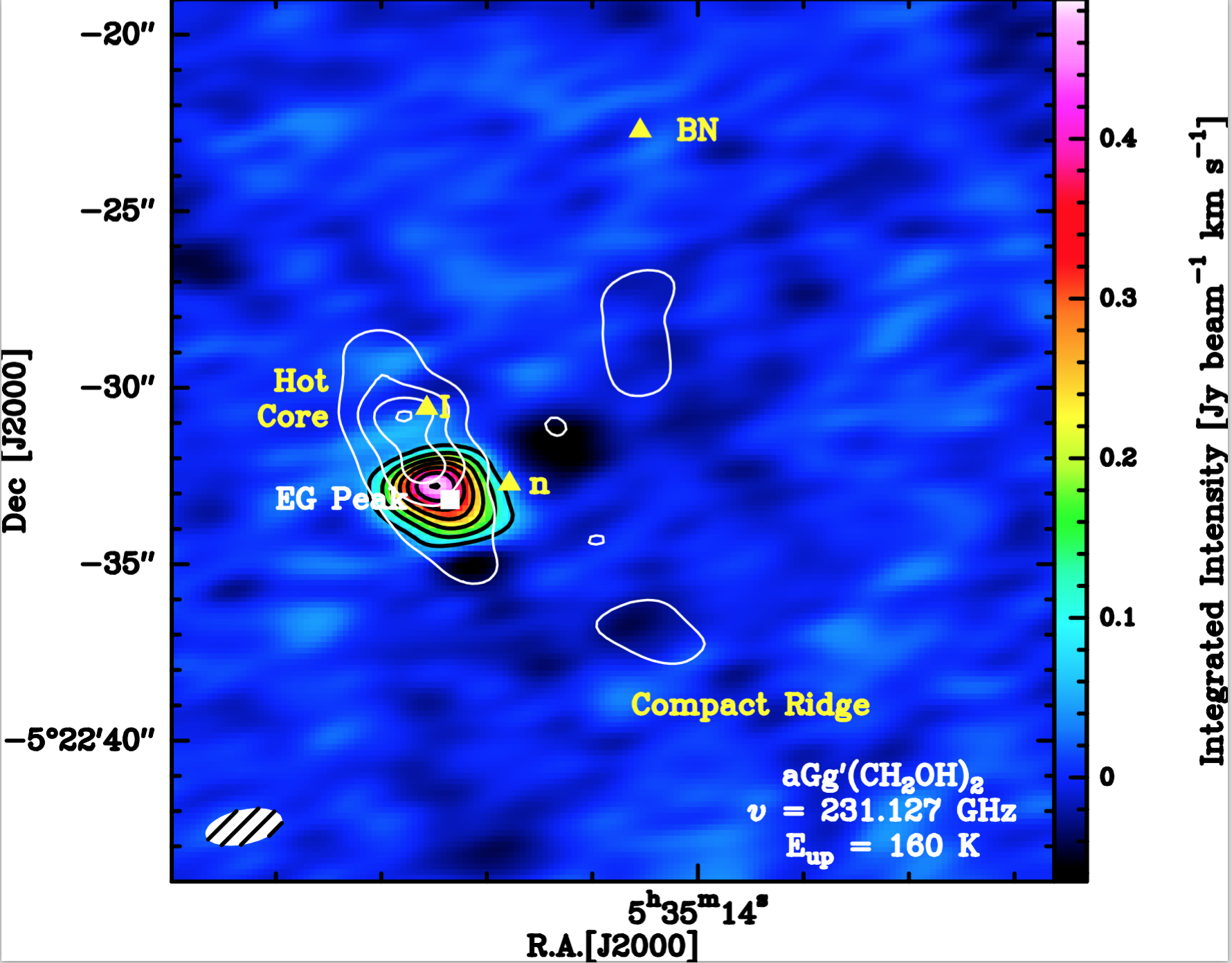}
\includegraphics[width=6cm]{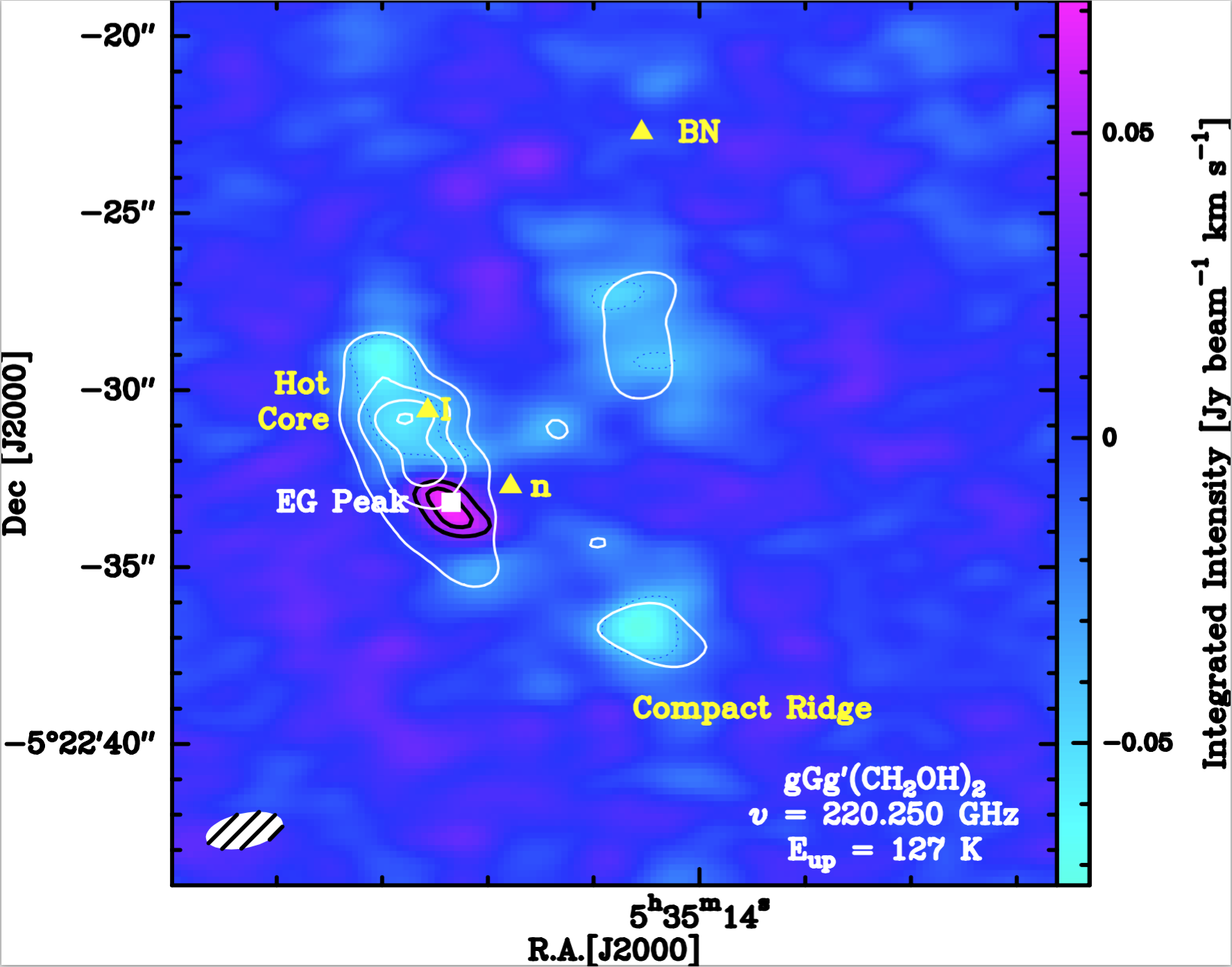}
\caption{{\it Left Panel:} CH$_3$COOH integrated emission map at 219016~MHz. The first contour and the level step are at 5$\sigma$  (where 1$\sigma$=9.3$\times$10$^{-3}$~Jy~beam$^{-1}$~km~s$^{-1}$). 
{\it Middle Panel:} aGg$^{\prime}$(CH$_2$OH)$_2$ integrated emission map at 231127~MHz. The first contour and the level step are at 5$\sigma$  (where 1$\sigma$=1.4$\times$10$^{-2}$~Jy~beam$^{-1}$~km~s$^{-1}$). 
{\it Right Panel:} gGg$^{\prime}$(CH$_2$OH)$_2$ integrated emission map at 220250~MHz. The contour levels are at -4, 4 and 6$\sigma$  (where 1$\sigma$=1$\times$10$^{-2}$~Jy~beam$^{-1}$~km~s$^{-1}$).
A narrow $v_\mathrm{LSR}$ interval (from 7 to 9~km/s) has been selected to reduce confusion by nearby lines (see Section~3.3 and Appendix~\ref{sec:app3}). 
Positions of the radio source I,  the BN object and the IR source n \citep[see][]{Goddi:2011} are indicated by yellow triangles.
The white square indicates the position of the Ethylene Glycol Peak ($\alpha_{J2000}$ = 05$^{h}$35$^{m}$14\fs47,  $\delta_{J2000}$ = -05$\degr$22$\arcmin$33$\farcs$17) by BD15. Finally, the continuum emission at 235~GHz is overlaid in white contours with a level step of 0.2~Jy~beam$^{-1}$ (Paper I).
}
\label{fg1}
\end{figure*}
%---------------

Our ALMA-Cycle 2 data and first results are given in a companion paper by \citet[hereafter Paper I]{Pagani:2017}. In this Letter, we focus on acetic acid (CH$_3$COOH) and the gGg$^{\prime}$ conformer of ethylene glycol (gGg$^{\prime}$(CH$_2$OH)$_2$) and report their first detection in Orion-KL. The detection of acetic acid in Orion-KL has not yet been reported, although a few transitions may be present in the IRAM 30m survey by \citet{Tercero:2011}. However, this species is known to be present in low-mass and high-mass star forming regions \citep[e.g.][]{Remijan:2003,Shiao:2010,Jorgensen:2016}. Regarding gGg$^{\prime}$-Ethylene glycol, this conformer has only been detected toward the Class 0 protostar IRAS 16293--2422  by \citet{Jorgensen:2016}. Incidentally, the most stable conformer of ethylene glycol (aGg$^{\prime}$) is detected towards low-mass, intermediate-mass and high-mass sources, including Orion-KL \citep[see e.g.][and references therein]{Fuente:2014,Lykke:2015,Brouillet:2015,Rivilla:2017}. 
In Section~\ref{sec:observations} we briefly describe our ALMA observations. Results and analysis are given and discussed in Sections~\ref{sec:datanalyse} and \ref{sec:discuss}, respectively.

%===============================================================
%
%-----------------------------------------------------------------------------------------------------------------------------
%-------------------------------------------- OBSERVATIONS -------------------------------------------------
%-----------------------------------------------------------------------------------------------------------------------------
\section{Observations and data reduction}
\label{sec:observations}
%-----------------------------------------------------------------------------------------------------------------------------
%
Acetic acid and ethylene glycol lines towards Orion--KL were observed with 37 antennas on 2014 December 29 and 39 antennas on 2014 December 30. The two following phase-tracking centers were used to perform the observations: $\alpha_{J2000}$ = 05$^{h}$35$^{m}$14\fs16,  $\delta_{J2000}$ = -05$\degr$22$\arcmin$31$\farcs$504 and $\alpha_{J2000}$ = 05$^{h}$35$^{m}$13\fs477,  $\delta_{J2000}$ = -05$\degr$22$\arcmin$08$\farcs$50. The observations lie in the frequency range 215.15 GHz to 252.04 GHz in band 6 and cover about 16~GHz of effective bandwidth with spectral resolution of about 0.7~km~s$^{-1}$. Data reduction and continuum subtraction were performed through the Common Astronomy Software Applications (CASA) software \citep{McMullin:2007}. The cleaning of the spectral lines was performed using the GILDAS software\footnote{http://www.iram.fr/IRAMFR/GILDAS/}. The resulting synthesized beam is typically 1.8$\arcsec$ $\times$ 1.1$\arcsec$ (P.A. of 84$\degr$). For further details, see Paper I.

%===============================================================
%
%-----------------------------------------------------------------------------------------------------------------------------
%--------------------------------------------DATA ANALYSIS-------------------------------------------------
%-----------------------------------------------------------------------------------------------------------------------------
\section{Data analysis and results}
\label{sec:datanalyse}
%-----------------------------------------------------------------------------------------------------------------------------
%
\subsection{Acetic acid and Ethylene glycol molecular frequencies}

We used the spectroscopic data parameters from \citet{Ilyushin:2008} and \citet{Ilyushin:2013} for acetic acid, with the following line selection criteria: Einstein spontaneous emission coefficient A$_\mathrm{ij}$ $\ge$ 5$\times$10$^{-5}$~s$^{-1}$ and upper level energy E$_\mathrm{up}$ $\le$ 400~K. For the partition function we adopt the complete rotational-torsional-vibrational partition function given by Calcutt, Woods, Carvajal et al. (to be submitted to MNRAS).

For both ethylene glycol conformers we used the spectroscopic data parameters from \citet{Christen:2003} and \citet{Muller:2004} that are available from the Cologne Database for Molecular Spectroscopy catalog \citep[CDMS,][]{Muller:2005}. More specifically, we searched for transitions up to E$_\mathrm{up}$ $\simeq$ 400~K, and A$_\mathrm{ij}$ $\ge$ 1$\times$10$^{-4}$~s$^{-1}$. The energy difference between the two conformers is about 200~cm$^{-1}$, the more stable conformer being the aGg$^{\prime}$-ethylene Glycol \citep{Muller:2004}. Further details about the difference between the aGg$^{\prime}$ and the gGg$^{\prime}$ conformer can be found in \citet[hereafter BD15]{Brouillet:2015}.

\subsection{LTE Modeling}
Our analysis is based on the assumption that local thermodynamic equilibrium (LTE) is reached. This assumption is reasonable given that LTE modeling of a thousand emissive transitions assigned to simple and complex molecules fits well the HIFI/Herschel observations performed towards  Orion-KL \citep[see][]{Crockett:2014}. In addition, we assume that all the species emit at the same rotational temperature within the same source size. We use the CLASS extension WEEDS \citep{Maret:2011} to model the acetic acid and ethylene glycol (both aGg$^{\prime}$ and gGg$^{\prime}$ conformer) emission, that we assume to be optically thin. We used the values derived for aGg$^{\prime}$(CH$_2$OH)$_2$ by BD15 as input parameter to initialize our models.

%++++++++++++++++
% TABLE 1
%++++++++++++++++
\begin{table*}[h!]
\caption{Best fit line parameters and derived peak column densities for acetic acid and ethylene glycol towards Orion--KL EGP.}
\label{tab1}
\centering
\begin{tabular}{lrrrr|rrrr} 
\hline\hline             
Molecule &  \multicolumn{4}{c}{Component 1} &  \multicolumn{4}{c}{Component 2}  \\
 & v (km~s$^{-1}$)& $\Delta$v$_{1/2}$ (km~s$^{-1}$)&T$\rm_{rot}$ (K)& N (10$^{15}$ cm$^{-2}$)& v (km~s$^{-1}$)& $\Delta$v$_{1/2}$ (km~s$^{-1}$)&T$\rm_{rot}$ (K)& N (10$^{15}$ cm$^{-2}$)\\
\hline
CH$_3$COOH & 7.9 & 2.5 & 140 & 12 & 5.1 & 2.3 & 140 & 3.3\\  
\hline
aGg$^{\prime}$(CH$_2$OH)$_2$ & 7.8 & 2.1& 140 & 6.8 & 5.1 & 2.1& 140 & 1.5 \\  
\hline  
gGg$^{\prime}$(CH$_2$OH)$_2$ & 7.8 & 2.1& 140 & 2.7 & 5.1 & 2.1& 140 & 0.66 \\  
\hline
\end{tabular}
\end{table*}

\subsection{Emission map}
The CH$_3$COOH, aGg$^{\prime}$(CH$_2$OH)$_2$ and gGg$^{\prime}$(CH$_2$OH)$_2$ emission maps integrated over the line profile are shown in Figure~\ref{fg1}. The nominal velocity of Orion--KL is $v_\mathrm{LSR}$=7.6~km~s$^{-1}$. It is important to note that the northwest extension seen in the acetic acid emission map is due to contamination by a U-line (see Appendix~\ref{sec:app3}) and is not related to the acetic acid emission. Indeed, although we use a restricted $v_\mathrm{LSR}$ interval to produce the maps, confusion still dominate the region (Paper I).

A salient result is that the distribution of the emission associated with these molecules is similar within the beam and the main emission peak is located about 2$^{\arcsec}$ southwest of the hot core, near both radio source I and IR source n.
This  peak corresponds to the "Ethylene Glycol Peak" (hereafter EGP) identified by BD15 for the aGg$^{\prime}$(CH$_2$OH)$_2$ conformer. An outstanding result is that, as for the aGg$^{\prime}$(CH$_2$OH)$_2$ molecule (BD15), the distribution of the emission associated with the acetic acid and  the gGg$^{\prime}$-ethylene glycol conformer differs from that of typical O-bearing species within Orion-KL. Indeed, the emission of the targeted  species appears to come from a compact source in the vicinity of the Hot Core region while the emission associated with O-bearing molecules, such as methyl formate \citep[e.g. see][and Appendix~\ref{sec:app4}]{Favre:2011}, is generally described by an extended V-shape within Orion-KL linking the Hot Core component to the Compact Ridge region and extending towards the BN object \citep[e.g.,][]{Guelin:2008}.

\subsection{Spectra}
\begin{figure}[h!]
\includegraphics[width=\hsize]{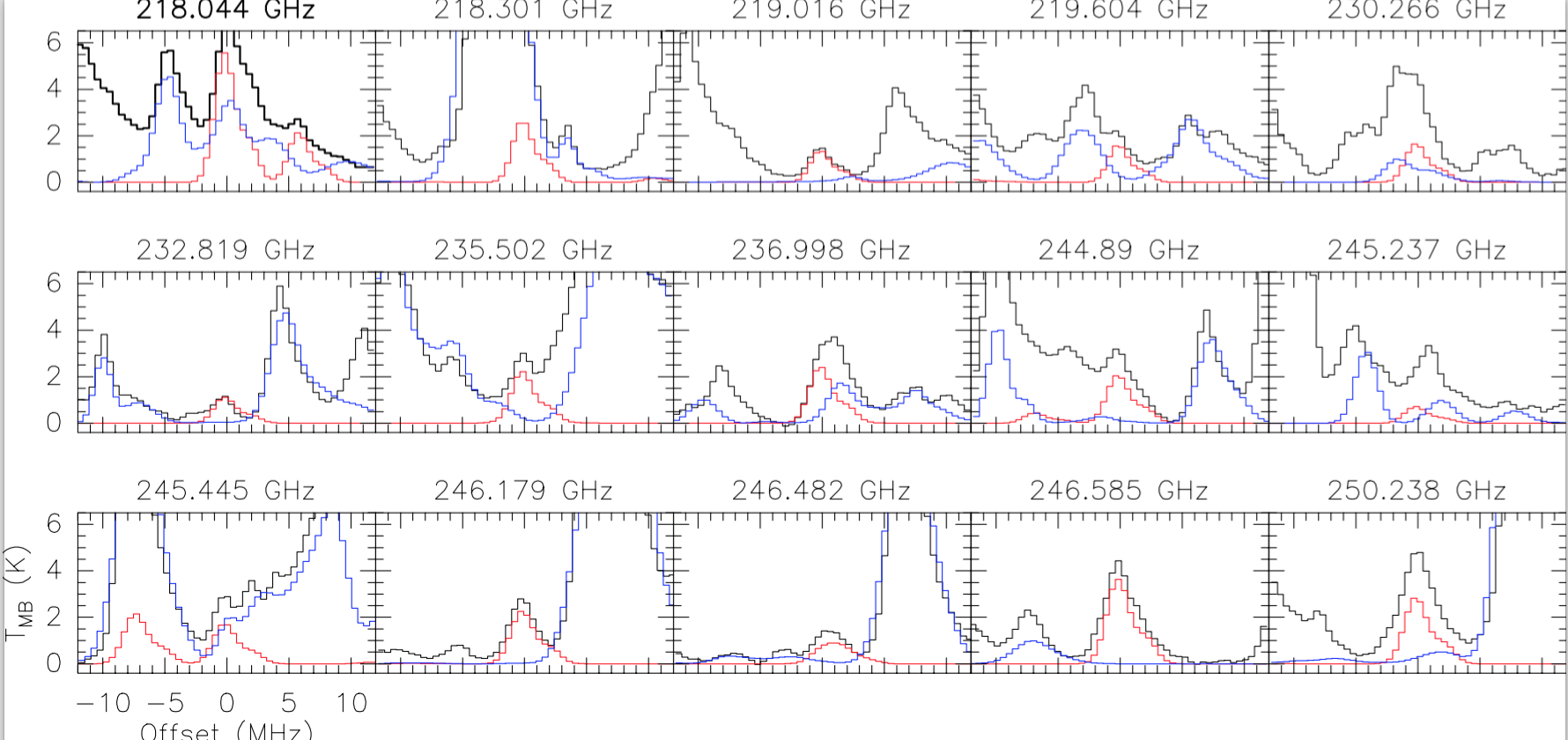}
\caption{ALMA observations (black) overlaid with the WEEDS model for acetic acid (red). The sum of the modeled emission from all the other species is overlaid in blue (Paper I).}
\label{fg2}
\end{figure}

\begin{figure}[h!]
\hspace*{0.25cm}\includegraphics[width=8.25cm]{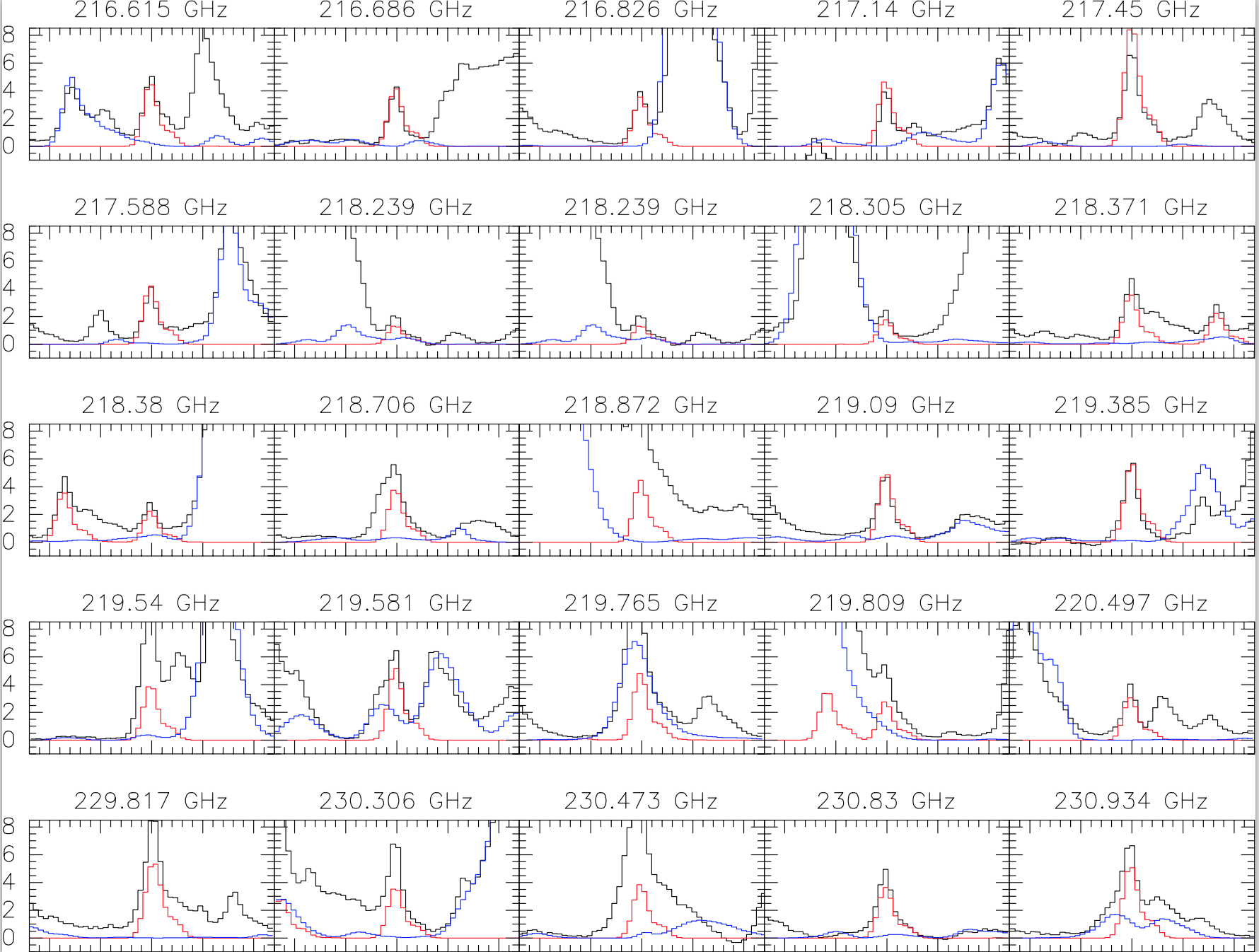}
\includegraphics[width=8.5cm]{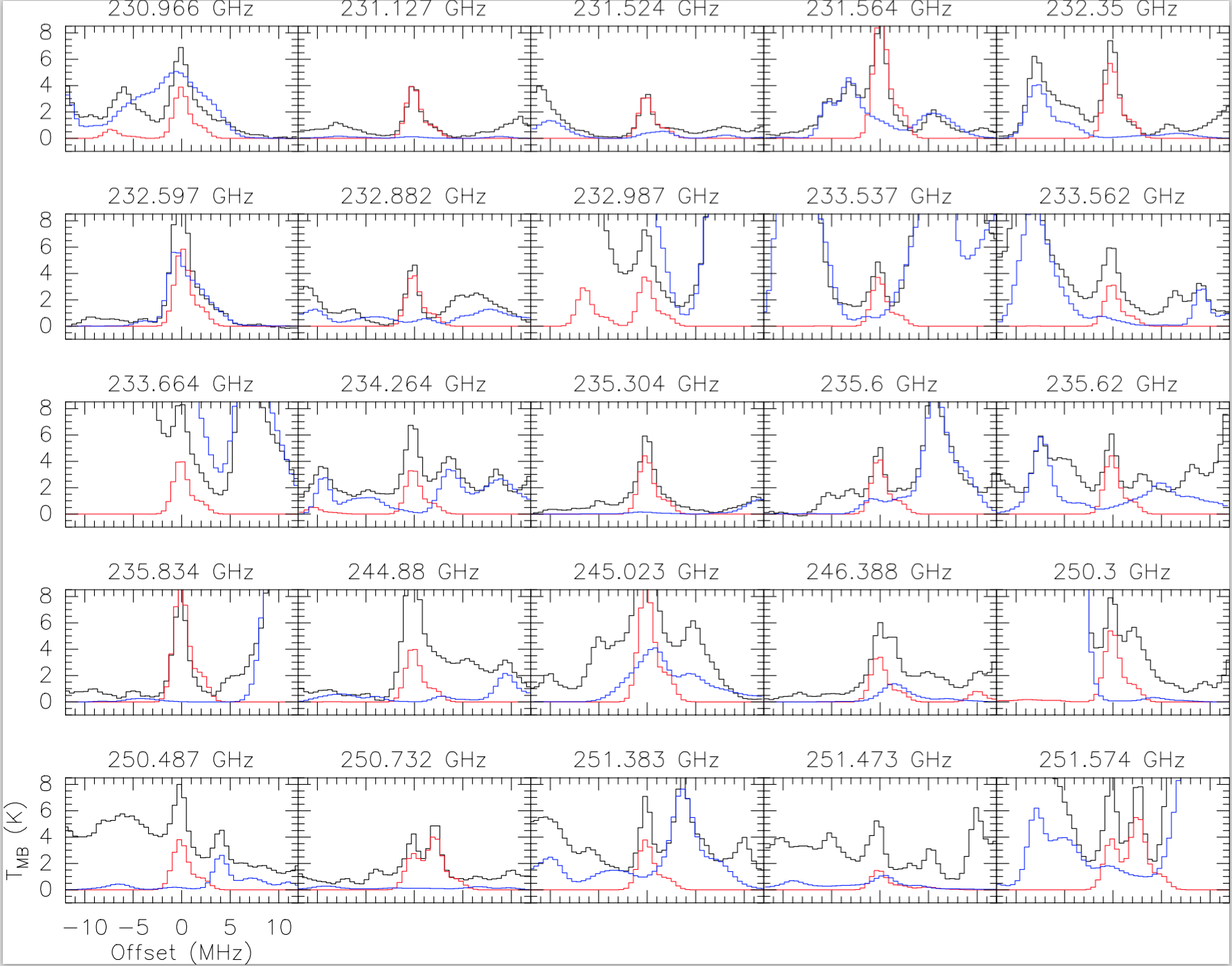}
\caption{ALMA observations (black) overlaid with the WEEDS model for aGg$^{\prime}$ ethylene glycol  (red). The sum of the modeled emission from all the other species is overlaid in blue (Paper I).}
\label{fg3}
\end{figure}

\begin{figure}[h!]
\includegraphics[width=\hsize]{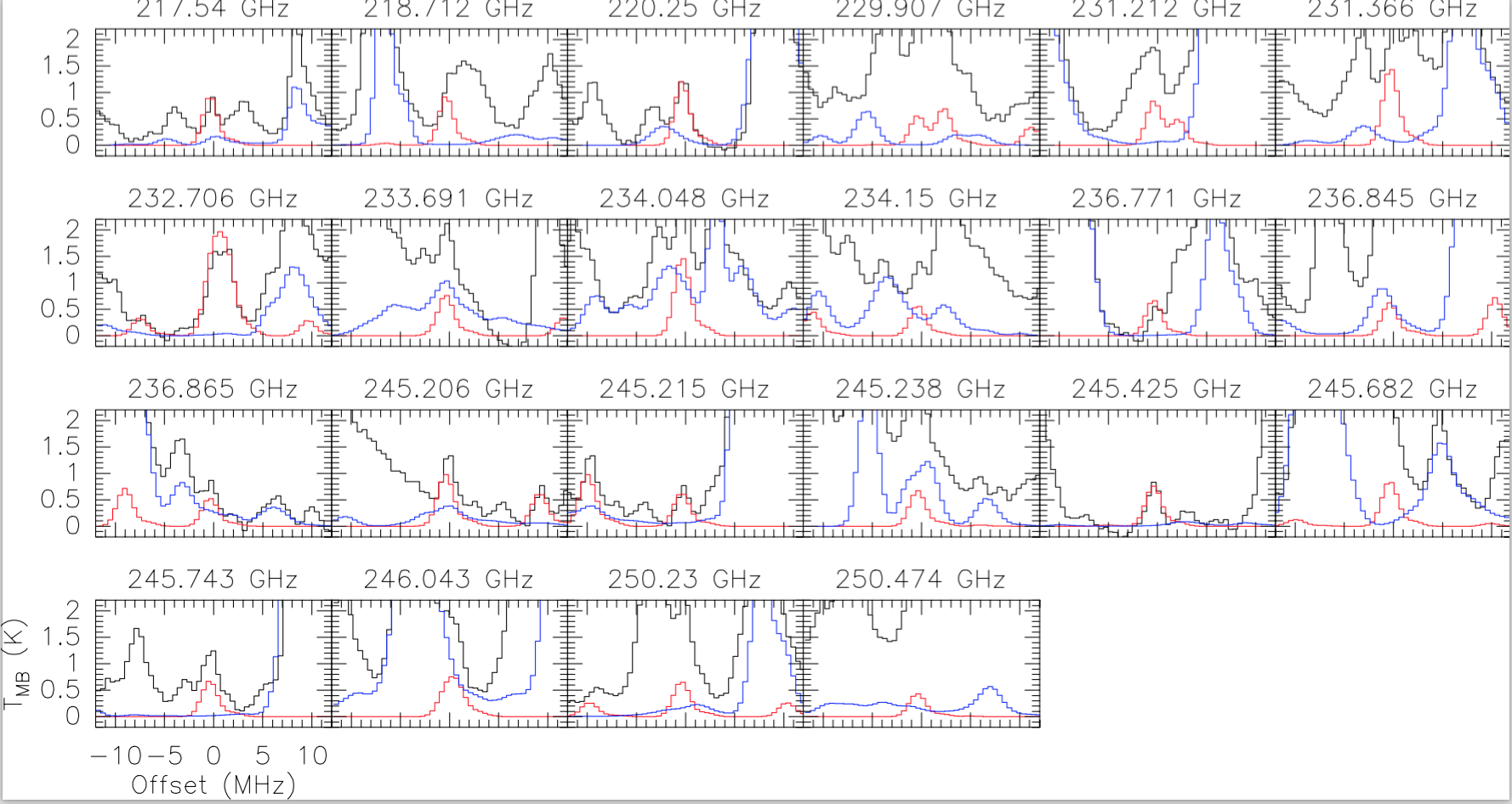}
\caption{ALMA observations (black) overlaid with the WEEDS model for gGg$^{\prime}$ ethylene glycol  (red). The sum of the modeled emission from all the other species is overlaid in blue (Paper I).}
\label{fg4}
\end{figure}

Spectra of a sample of the most intense transitions (i.e. emitting above the 5$\sigma$ level) of acetic acid (15 transitions from E$\rm_{up}$=70~K up to 318~K, including 5 unblended), aGg$^{\prime}$ ethylene glycol (50 transitions from E$\rm_{up}$=111~K up to 266~K, including 19 unblended) and gGg$^{\prime}$ ethylene glycol (22 transitions from E$\rm_{up}$=102~K up to 216~K, including 5 unblended) towards the EGP region are displayed in Figs~\ref{fg2}, \ref{fg3} and \ref{fg4}, respectively. In addition, our best "by eye" WEEDS fits together with the sum of the modeled emission from all the other species in the region (Paper I) are also overlaid in these figures. Tables~\ref{table:B1}, \ref{table:B2} and \ref{table:B3} in Appendix~\ref{sec:app2}, list the spectroscopic line parameters for the displayed acetic acid, aGg$^{\prime}$--ethylene glycol and gGg$^{\prime}$--ethylene glycol transitions, respectively.
The bulk of the emission associated with the targeted molecules peaks at about 7.8--7.9~km~s$^{-1}$. Nonetheless, all the line profiles display an extended blue-shifted wing. Thus, two velocity components, one around 8~km~s$^{-1}$ and the other one at about 5~km~s$^{-1}$, are required to fit the emission. 
The model parameters that best reproduce the ALMA observations of acetic acid and ethylene glycol (both conformers) in the direction of the EGP region are summarized in Table~\ref{tab1}. In the present analysis we assume an overall uncertainty in the range 30$\%$--40$\%$. 

%------------------------------------------------------------------
% --- FIGURE 2 ---
%-----------------------------------------------------------------
%---------------

%------------------------------------------------------------------
% --- FIGURE 3 ---
%-----------------------------------------------------------------
%---------------

%------------------------------------------------------------------
% --- FIGURE 4 ---
%-----------------------------------------------------------------
%---------------

\subsection{Column densities and Relative abundances}
Table~\ref{tab1} gives the derived CH$_3$COOH, aGg$^{\prime}$(CH$_2$OH)$_2$ and gGg$^{\prime}$(CH$_2$OH)$_2$ peak column densities assuming a source size of 3$\arcsec$ for each velocity component. We note that our best aGg$^{\prime}$(CH$_2$OH)$_2$ fit result (v, $\Delta_{v}$, T$\rm_{rot}$ and N) is consistent within the uncertainties ($\sim$30$\%$-40$\%$) with the value reported by BD15.

Table~\ref{tab2} lists the relative abundance ratios for acetic acid and ethylene glycol derived from our best model results (see Table~\ref{tab1}) towards both velocity components observed in direction of the EGP peak. The derived abundance ratios are equal within the error bars for both velocity components. 
It is important to note that BD15 reported an upper limit on the aGg$^{\prime}$(CH$_2$OH)$_2$/gGg$^{\prime}$(CH$_2$OH)$_2$ ratio of 5. This discrepancy apparently results from an underestimate of the limit on the gGg$^{\prime}$(CH$_2$OH)$_2$ column density by BD15.

%++++++++++++++++
% TABLE 2
%++++++++++++++++
\begin{table}[h!]
\caption{Relative abundances.}\label{tab2}
\centering
\begin{tabular}{lccc} 
\hline\hline             
Component & \large $\frac{\rm CH_{3}COOH}{\rm aGg^{\prime}(CH_2OH)_2}$ & \large $\frac{\rm CH_{3}COOH}{\rm gGg^{\prime}(CH_2OH)_2}$ & \large $\frac{\rm aGg^{\prime}(CH_2OH)_2}{\rm gGg^{\prime}(CH_2OH)_2}$  \\
\hline
{8~km~s$^{-1}$}& 1.8 & 4.4 & 2.5 \\
\hline
{5~km~s$^{-1}$}& 2.4 & 5.0 & 2.3 \\
\hline
\end{tabular}
\tablefoot{The given values have uncertainties of 40$\%$--50$\%$.}
\end{table}
%===============================================================
%
%-----------------------------------------------------------------------------------------------------------------------------
%---------------------------------------DISCUSSION -------------------------------------------------
%-----------------------------------------------------------------------------------------------------------------------------
\section{Discussion}
\label{sec:discuss}
\begin{figure}[h!]
\centering
\includegraphics[width=6.7cm]{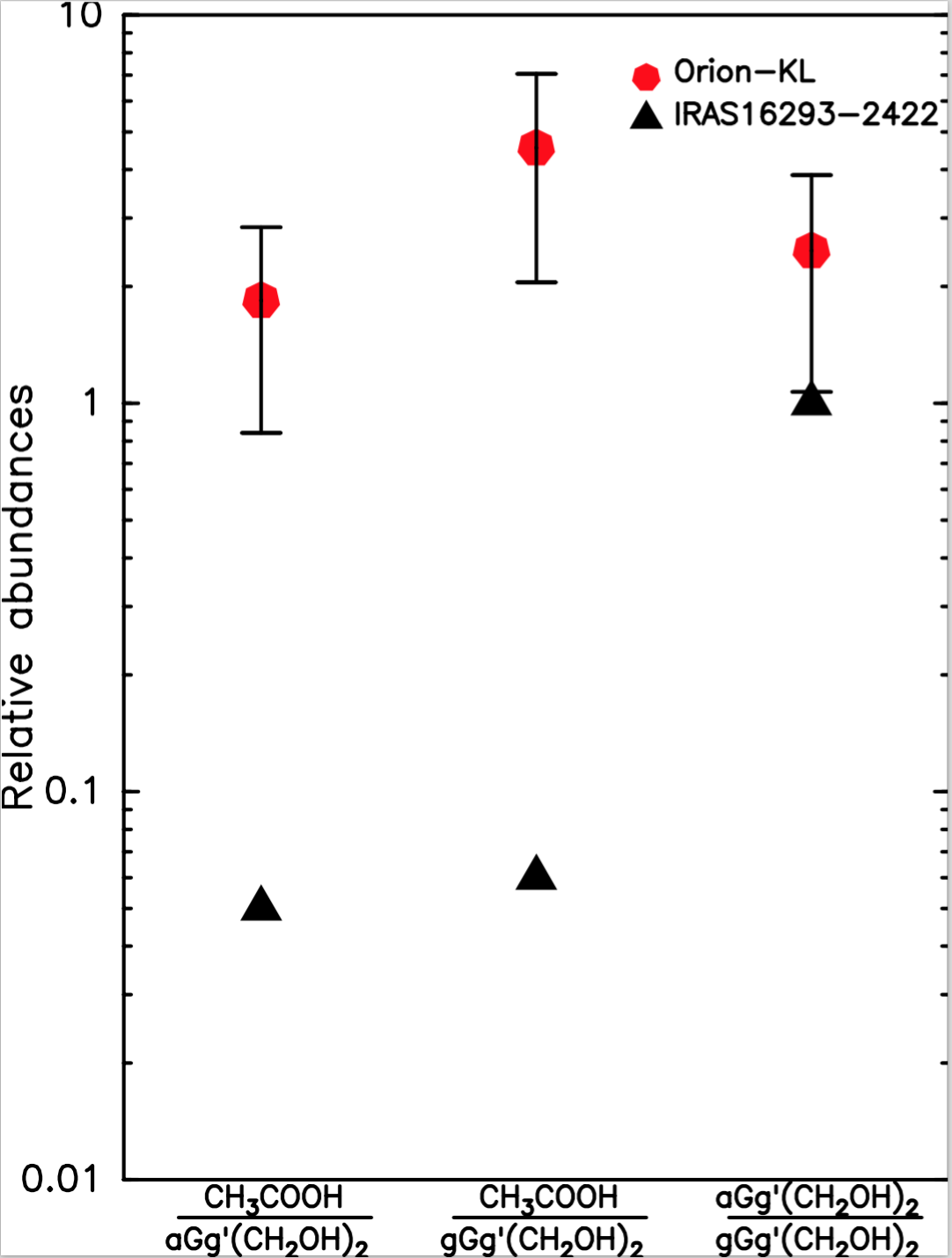}
\caption{Acetic acid and ethylene glycol abundance ratios towards Orion-KL (red circles, this study) and IRAS 16293-2422 \citep[black triangles,][]{Jorgensen:2016}.
The ratios for Orion-KL are obtained from the sum of the velocity components given in Table~\ref{tab1}.
}
\label{fg5}
\end{figure}

%-----------------------------------------------------------------------------------------------------------------------------
In Figure~\ref{fg5}, we show the relative abundance ratios, CH$_3$COOH:aGg$^{\prime}$(CH$_2$OH)$_2$:gGg$^{\prime}$(CH$_2$OH)$_2$,  derived in this study along with the ones derived towards the low-mass protostar IRAS 16293--2422  by \citet{Jorgensen:2016}.
It is immediately apparent that the CH$_3$COOH:(CH$_2$OH)$_2$ ratios measured in direction of Orion-KL are larger than those of the low-mass protostar IRAS 16293--2422 by over an order of magnitude. Also, we note that the aGg$^{\prime}$(CH$_2$OH)$_2$:gGg$^{\prime}$(CH$_2$OH)$_2$ ratio seems to be, within the error bars, the same for both regions. The fact that \citet{Jorgensen:2016} assumed different rotational temperatures for the two conformers to estimate this ratio might explain the slight difference. \citet{Lykke:2015} have shown that the source luminosities are likely correlated with relative abundances of complex organic molecules.
These findings lead one to ask whether and how the physical conditions in these regions, in particular Orion-KL, impact the production and the possible release to the gas-phase, of these species. 

Both CH$_3$COOH and (CH$_2$OH)$_2$ are believed to mainly be formed on the icy-surface of grains, although gas-phase formation routes cannot be ruled out \citep[see e.g.][]{Laas:2011,Rivilla:2017}. Interestingly enough, \citet{Garrod:2008} have shown that ethylene glycol is produced more efficiently in grain mantles in comparison to acetic acid by at least one order of magnitude. This naturally explains the observation that the abundance ratio CH$_3$COOH/(CH$_2$OH)$_2$ is lower in low-mass star forming regions. However, regarding Orion-KL, an additional mechanism is required to explain the over abundance of CH$_3$COOH. It is noteworthy that \citet[][]{Wright:2017} have recently proposed that a bullet of matter ejected during the explosive event that occurred $\sim$500-700 years ago \citep[][]{Nissen:2012} has impacted the EGP region. More specifically, using high angular resolution ALMA observations, \citet[][]{Wright:2017} have reported the presence of a molecular ring in HC$_3$N, HCN and SO$_2$ which is not associated with continuum emission. In that context, it is important to note that the distribution of the acetic acid and ethylene glycol is co-spatial with this ring (Figure~\ref{fgE} in Appendix~\ref{sec:app5}). In addition, both acetic acid and ethylene glycol line profiles present a blue-shifted emission wing (i.e. the 5 km~s$^{-1}$ velocity component), this specific asymmetric line profile being also observed for other molecules in this region (e.g. methanol, formic acid, Paper I). 
These findings strongly suggest that this region is peculiar and is different from other star-forming regions. Indeed, the impact that took place here has led to the release of icy COMs in the gas-phase, generating the observed gas motions together with a rich molecular composition that may reflect gas-phase chemistry in an induced shock/post-shock stage.

%------------------------------------------------------------------
% --- FIGURE 5 ---
%-----------------------------------------------------------------
%---------------

%===============================================================
%
%-----------------------------------------------------------------------------------------------------------------------------
%-----------------------------------------ACKNOWLEDGEMENTS-----------------------------------
%-----------------------------------------------------------------------------------------------------------------------------
%
\begin{acknowledgements}
See Appendix~\ref{sec:app1}.
\end{acknowledgements}

%===============================================================
%
%-----------------------------------------------------------------------------------------------------------------------------
%-----------------------------------------BIBLIO----------------------------------
%-----------------------------------------------------------------------------------------------------------------------------
%
\bibliographystyle{aa}
%\bibliography{/Users/cecilefavre/Documents/articles/biblio}

%
%===============================================================
%

%===============================================================
%
%-----------------------------------------------------------------------------------------------------------------------------
%-----------------------------------------APPENDIX----------------------------------
%-----------------------------------------------------------------------------------------------------------------------------
%
\begin{appendix}

%------------------------------------------------------------------
% ---Acknowledgements ---
%-----------------------------------------------------------------
\section{Acknowledgements }
\label{sec:app1}
CF acknowledges support from the Italian Ministry of Education, Universities and Research, project SIR (RBSI14ZRHR). The authors thank Hannah Calcutt for providing the acetic acid partition function. We also thanks Melvyn Wright and Rick Plambeck for their HC$_3$N emission map. This work was carried out in part at the Jet Propulsion Laboratory, which is operated for NASA by the California Institute of Technology. MC acknowledges financial funding from the project FIS2014-53448-C2-2-P (MINECO, Spain), and CMST COST Action CM1405 MOLIM. CF and MC acknowledge support from CMST COST Action CM1401 Our Astro-Chemical History. LP thanks Arnaud Belloche and H.S.P. M\"uller for their help with molecular spectroscopy data. CF thanks Claudio Codella for enlightening discussion on shocks. Finally, CF thanks Vianney Taquet and Franck Hersant for fruitful discussion about the binding energies and acetic acid formation routes.

%------------------------------------------------------------------
% ---Spectroscopic line parameters ---
%-----------------------------------------------------------------
\section{Spectroscopic line parameters}
\label{sec:app2}

Tables~\ref{table:B1}, \ref{table:B2} and \ref{table:B3} list the spectroscopic line parameters for the acetic acid, aGg$^{\prime}$--ethylene glycol and gGg$^{\prime}$--ethylene glycol lines that are displayed in Figures~\ref{fg2}, \ref{fg3} and \ref{fg4}, respectively.

\begin{table*}[h!]
\caption{Spectroscopic data of the acetic acid lines displayed in Fig. \ref{fg2}}       
\label{table:B1}      
\centering          
\begin{tabular}{c cc c c c c }     % 6 columns 
\hline\hline       
Frequency&Symmetry&\multicolumn{2}{c}{Quantum number}&A$_{u,l}$&g$_{up}$&E$_{up}$\\
(MHz)&&J$_{K_a,K_c}$,\emph{v$_t$}(up)&J$_{K_a,K_c}$,\emph{v$_t$}(low)&(s$^{-1}$)&&(K)\\
\hline                    
218044.2146 & A & 20$_{(0,20)}$,\emph{v$_t$}=0 & 19$_{(1,19)}$,\emph{v$_t$}=0 & 2.25e-05 & 82 & 112.2\\
218044.2146 & A & 20$_{(1,20)}$,\emph{v$_t$}=0 & 19$_{(0,19)}$,\emph{v$_t$}=0 & 2.25e-05 & 82 & 112.2\\
218044.2146 & A & 20$_{(0,20)}$,\emph{v$_t$}=0 & 19$_{(0,19)}$,\emph{v$_t$}=0 & 6.06e-05 & 82 & 112.2\\
218044.2146 & A & 20$_{(1,20)}$,\emph{v$_t$}=0 & 19$_{(1,19)}$,\emph{v$_t$}=0 & 6.06e-05 & 82 & 112.2\\
218301.0685 & E & 20$_{(0,20)}$,\emph{v$_t$}=1 & 19$_{(0,19)}$,\emph{v$_t$}=1 & 2.44e-08 & 82 & 217.4\\
218301.0685 & E & 20$_{(1,20)}$,\emph{v$_t$}=1 & 19$_{(1,19)}$,\emph{v$_t$}=1 & 2.44e-08 & 82 & 217.4\\
218301.0685 & E & 20$_{(0,20)}$,\emph{v$_t$}=1 & 19$_{(1,19)}$,\emph{v$_t$}=1 & 8.31e-05 & 82 & 217.4\\
218301.0685 & E & 20$_{(1,20)}$,\emph{v$_t$}=1 & 19$_{(0,19)}$,\emph{v$_t$}=1 & 8.31e-05 & 82 & 217.4\\
219016.0364 & E & 20$_{(0,20)}$,\emph{v$_t$}=2 & 19$_{(1,19)}$,\emph{v$_t$}=2 & 6.41e-05 & 82 & 309.0\\
219016.0365 & E & 20$_{(0,20)}$,\emph{v$_t$}=2 & 19$_{(0,19)}$,\emph{v$_t$}=2 & 1.77e-05 & 82 & 309.0\\
219016.0385 & E & 20$_{(1,20)}$,\emph{v$_t$}=2 & 19$_{(1,19)}$,\emph{v$_t$}=2 & 1.77e-05 & 82 & 309.0\\
219016.0386 & E & 20$_{(1,20)}$,\emph{v$_t$}=2 & 19$_{(0,19)}$,\emph{v$_t$}=2 & 6.41e-05 & 82 & 309.0\\
219603.9437 & A & 20$_{(0,20)}$,\emph{v$_t$}=2 & 19$_{(1,19)}$,\emph{v$_t$}=2 & 6.51e-06 & 82 & 289.1\\
219603.9437 & A & 20$_{(1,20)}$,\emph{v$_t$}=2 & 19$_{(1,19)}$,\emph{v$_t$}=2 & 7.77e-05 & 82 & 289.1\\
219603.9446 & A & 20$_{(1,20)}$,\emph{v$_t$}=2 & 19$_{(0,19)}$,\emph{v$_t$}=2 & 6.51e-06 & 82 & 289.1\\
219603.9446 & A & 20$_{(0,20)}$,\emph{v$_t$}=2 & 19$_{(0,19)}$,\emph{v$_t$}=2 & 7.77e-05 & 82 & 289.1\\
230266.0061 & A & 21$_{(0,21)}$,\emph{v$_t$}=2 & 20$_{(1,20)}$,\emph{v$_t$}=2 & 2.71e-05 & 86 & 300.2\\
230266.0061 & A & 21$_{(1,21)}$,\emph{v$_t$}=2 & 20$_{(0,20)}$,\emph{v$_t$}=2 & 2.71e-05 & 86 & 300.2\\
230266.0061 & A & 21$_{(0,21)}$,\emph{v$_t$}=2 & 20$_{(0,20)}$,\emph{v$_t$}=2 & 7.02e-05 & 86 & 300.2\\
230266.0061 & A & 21$_{(1,21)}$,\emph{v$_t$}=2 & 20$_{(1,20)}$,\emph{v$_t$}=2 & 7.02e-05 & 86 & 300.2\\
232818.7007 & E & 19$_{(2,17)}$,\emph{v$_t$}=2 & 18$_{(3,16)}$,\emph{v$_t$}=2 & 7.31e-05 & 78 & 317.7\\
232818.7077 & E & 19$_{(3,17)}$,\emph{v$_t$}=2 & 18$_{(3,16)}$,\emph{v$_t$}=2 & 1.06e-05 & 78 & 317.7\\
232818.7318 & E & 19$_{(2,17)}$,\emph{v$_t$}=2 & 18$_{(2,16)}$,\emph{v$_t$}=2 & 1.06e-05 & 78 & 317.7\\
232818.7388 & E & 19$_{(3,17)}$,\emph{v$_t$}=2 & 18$_{(2,16)}$,\emph{v$_t$}=2 & 7.31e-05 & 78 & 317.7\\
235501.6832 & A & 20$_{(2,18)}$,\emph{v$_t$}=1 & 19$_{(3,17)}$,\emph{v$_t$}=1 & 3.21e-05 & 82 & 241.5\\
235501.6832 & A & 20$_{(3,18)}$,\emph{v$_t$}=1 & 19$_{(2,17)}$,\emph{v$_t$}=1 & 3.21e-05 & 82 & 241.5\\
235501.6832 & A & 20$_{(2,18)}$,\emph{v$_t$}=1 & 19$_{(2,17)}$,\emph{v$_t$}=1 & 6.14e-05 & 82 & 241.5\\
235501.6832 & A & 20$_{(3,18)}$,\emph{v$_t$}=1 & 19$_{(3,17)}$,\emph{v$_t$}=1 & 6.14e-05 & 82 & 241.5\\
236998.1508 & A & 21$_{(1,20)}$,\emph{v$_t$}=1 & 20$_{(1,19)}$,\emph{v$_t$}=1 & 1.01e-04 & 86 & 245.0\\
236998.1508 & A & 21$_{(2,20)}$,\emph{v$_t$}=1 & 20$_{(2,19)}$,\emph{v$_t$}=1 & 1.01e-04 & 86 & 245.0\\
236998.1508 & A & 21$_{(1,20)}$,\emph{v$_t$}=1 & 20$_{(2,19)}$,\emph{v$_t$}=1 & 1.42e-07 & 86 & 245.0\\
236998.1508 & A & 21$_{(2,20)}$,\emph{v$_t$}=1 & 20$_{(1,19)}$,\emph{v$_t$}=1 & 1.42e-07 & 86 & 245.0\\
244889.6209 & A & 20$_{(3,17)}$,\emph{v$_t$}=1 & 19$_{(3,16)}$,\emph{v$_t$}=1 & 2.42e-05 & 82 & 249.1\\
244889.6209 & A & 20$_{(4,17)}$,\emph{v$_t$}=1 & 19$_{(4,16)}$,\emph{v$_t$}=1 & 2.42e-05 & 82 & 249.1\\
244889.6209 & A & 20$_{(3,17)}$,\emph{v$_t$}=1 & 19$_{(4,16)}$,\emph{v$_t$}=1 & 7.48e-05 & 82 & 249.1\\
244889.6209 & A & 20$_{(4,17)}$,\emph{v$_t$}=1 & 19$_{(3,16)}$,\emph{v$_t$}=1 & 7.48e-05 & 82 & 249.1\\
245237.0813 & A & 12$_{(11,1)}$,\emph{v$_t$}=1 & 11$_{(10,2)}$,\emph{v$_t$}=1 & 7.14e-05 & 50 & 187.2\\
245444.9402 & E & 11$_{(11,1)}$,\emph{v$_t$}=0 & 10$_{(10,1)}$,\emph{v$_t$}=0 & 8.23e-05 & 46 & 70.1\\
246179.2041 & A & 21$_{(2,19)}$,\emph{v$_t$}=1 & 20$_{(3,18)}$,\emph{v$_t$}=1 & 3.88e-05 & 86 & 253.3\\
246179.2041 & A & 21$_{(3,19)}$,\emph{v$_t$}=1 & 20$_{(2,18)}$,\emph{v$_t$}=1 & 3.88e-05 & 86 & 253.3\\
246179.2041 & A & 21$_{(2,19)}$,\emph{v$_t$}=1 & 20$_{(2,18)}$,\emph{v$_t$}=1 & 6.89e-05 & 86 & 253.3\\
246179.2041 & A & 21$_{(3,19)}$,\emph{v$_t$}=1 & 20$_{(3,18)}$,\emph{v$_t$}=1 & 6.89e-05 & 86 & 253.3\\
246481.9960 & A & 19$_{(3,16)}$,\emph{v$_t$}=2 & 18$_{(4,15)}$,\emph{v$_t$}=2 & 6.01e-05 & 78 & 309.9\\
246584.8477 & E & 18$_{(5,13)}$,\emph{v$_t$}=0 & 17$_{(6,12)}$,\emph{v$_t$}=0 & 6.11e-05 & 74 & 129.1\\
246584.8511 & E & 18$_{(6,13)}$,\emph{v$_t$}=0 & 17$_{(6,12)}$,\emph{v$_t$}=0 & 2.23e-05 & 74 & 129.1\\
246584.8724 & E & 18$_{(5,13)}$,\emph{v$_t$}=0 & 17$_{(5,12)}$,\emph{v$_t$}=0 & 2.23e-05 & 74 & 129.1\\
246584.8759 & E & 18$_{(6,13)}$,\emph{v$_t$}=0 & 17$_{(5,12)}$,\emph{v$_t$}=0 & 6.11e-05 & 74 & 129.1\\
250237.9675 & E & 23$_{(0,23)}$,\emph{v$_t$}=1 & 22$_{(0,22)}$,\emph{v$_t$}=1 & 1.11e-04 & 94 & 251.9\\
250237.9675 & E & 23$_{(1,23)}$,\emph{v$_t$}=1 & 22$_{(1,22)}$,\emph{v$_t$}=1 & 1.11e-04 & 94 & 251.9\\
250237.9675 & E & 23$_{(0,23)}$,\emph{v$_t$}=1 & 22$_{(1,22)}$,\emph{v$_t$}=1 & 1.54e-05 & 94 & 251.9\\
250237.9675 & E & 23$_{(1,23)}$,\emph{v$_t$}=1 & 22$_{(0,22)}$,\emph{v$_t$}=1 & 1.54e-05 & 94 & 251.9\\
\hline                  
\end{tabular}
\end{table*}

\longtab[2]{
\begin{longtable}{c c c c c c}
\caption{\label{table:B2}Spectroscopic data of the aGg' ethylene glycol lines displayed in Fig. \ref{fg3}}\\
\hline
\hline
Frequency&\multicolumn{2}{c}{Quantum number\tablefootmark{a} }&A$_{u,l}$&g$_{up}$&E$_{up}$\\
(MHz)&J$_{K_a,K_c}$,\emph{v}(up)&J$_{K_a,K_c}$,\emph{v}(low)&s$^{-1}$&&(K)\\
\hline
\endfirsthead
\caption{Continued.} \\
\hline
Frequency&\multicolumn{2}{c}{Quantum number\tablefootmark{a} }&A$_{u,l}$&g$_{up}$&E$_{up}$\\
(MHz)&J$_{K_a,K_c}$,\emph{v}(up)&J$_{K_a,K_c}$,\emph{v}(low)&(s$^{-1}$)&&(K)\\
\hline
\endhead
\hline
\endfoot
\hline
\endlastfoot
216614.952  &  20$_{(3,17)}$,\emph{v}=1 & 19$_{(3,16)}$,\emph{v}=0 & 2.222E-04 & 369 & 110.8\\
216685.815  &  21$_{(3,19)}$,\emph{v}=1 & 20$_{(3,18)}$,\emph{v}=0 & 2.030E-04 & 387 & 117.3\\
216826.112  &  20$_{(5,15)}$,\emph{v}=1 & 19$_{(5,14)}$,\emph{v}=0 & 1.792E-04 & 369 & 116.8\\
217139.723  &  21$_{(4,17)}$,\emph{v}=0 & 20$_{(4,16)}$,\emph{v}=1 & 2.423E-04 & 387 & 123.9\\
217449.995  &  24$_{(1,24)}$,\emph{v}=0 & 23$_{(1,23)}$,\emph{v}=1 & 2.520E-04 & 441 & 136.4\\
217450.270  &  24$_{(0,24)}$,\emph{v}=0 & 23$_{(0,23)}$,\emph{v}=1 & 2.520E-04 & 343 & 136.4\\
217587.548  &  21$_{(2,19)}$,\emph{v}=1 & 20$_{(2,18)}$,\emph{v}=0 & 2.654E-04 & 301 & 117.2\\
218238.988  &  22$_{(17,5)}$,\emph{v}=0 & 21$_{(17,4)}$,\emph{v}=1 & 1.018E-04 & 315 & 266.2\\
218238.988  &  22$_{(17,6)}$,\emph{v}=0 & 21$_{(17,5)}$,\emph{v}=1 & 1.018E-04 & 405 & 266.2\\
218304.671  &  22$_{(16,6)}$,\emph{v}=0 & 21$_{(16,5)}$,\emph{v}=1 & 1.192E-04 & 315 & 250.0\\
218304.671  &  22$_{(16,7)}$,\emph{v}=0 & 21$_{(16,6)}$,\emph{v}=1 & 1.192E-04 & 405 & 250.0\\
218371.495  &  22$_{(4,19)}$,\emph{v}=0 & 21$_{(4,18)}$,\emph{v}=1 & 1.881E-04 & 405 & 132.6\\
218379.983  &  22$_{(15,7)}$,\emph{v}=0 & 21$_{(15,6)}$,\emph{v}=1 & 1.355E-04 & 315 & 234.8\\
218379.983  &  22$_{(15,8)}$,\emph{v}=0 & 21$_{(15,7)}$,\emph{v}=1 & 1.355E-04 & 405 & 234.8\\
218705.810  &  22$_{(12,10)}$,\emph{v}=0 & 21$_{(12,9)}$,\emph{v}=1 & 1.786E-04 & 315 & 195.1\\
218705.810  &  22$_{(12,11)}$,\emph{v}=0 & 21$_{(12,10)}$,\emph{v}=1 & 1.786E-04 & 405 & 195.1\\
218872.112  &  22$_{(11,12)}$,\emph{v}=0 & 21$_{(11,11)}$,\emph{v}=1 & 1.911E-04 & 405 & 183.8\\
218872.112  &  22$_{(11,11)}$,\emph{v}=0 & 21$_{(11,10)}$,\emph{v}=1 & 1.911E-04 & 315 & 183.8\\
219089.720  &  22$_{(10,13)}$,\emph{v}=0 & 21$_{(10,12)}$,\emph{v}=1 & 2.027E-04 & 405 & 173.5\\
219089.728  &  22$_{(10,12)}$,\emph{v}=0 & 21$_{(10,11)}$,\emph{v}=1 & 2.027E-04 & 315 & 173.5\\
219385.178  &  22$_{(9,14)}$,\emph{v}=0 & 21$_{(9,13)}$,\emph{v}=1 & 2.136E-04 & 405 & 164.3\\
219385.426  &  22$_{(9,13)}$,\emph{v}=0 & 21$_{(9,12)}$,\emph{v}=1 & 2.136E-04 & 315 & 164.3\\
219540.443  &  22$_{(2,21)}$,\emph{v}=1 & 21$_{(2,20)}$,\emph{v}=0 & 2.539E-04 & 315 & 122.1\\
219580.672  &  22$_{(1,21)}$,\emph{v}=1 & 21$_{(1,20)}$,\emph{v}=0 & 2.568E-04 & 405 & 122.0\\
219764.926  &  20$_{(4,16)}$,\emph{v}=1 & 19$_{(4,15)}$,\emph{v}=0 & 2.454E-04 & 369 & 113.5\\
219809.406  &  22$_{(8,14)}$,\emph{v}=0 & 21$_{(8,13)}$,\emph{v}=1 & 2.238E-04 & 315 & 156.0\\
220496.592  &  22$_{(7,15)}$,\emph{v}=0 & 21$_{(7,14)}$,\emph{v}=1 & 2.339E-04 & 315 & 148.8\\
229816.573  &  23$_{(9,15)}$,\emph{v}=0 & 22$_{(9,14)}$,\emph{v}=1 & 2.499E-04 & 329 & 175.6\\
229817.111  &  23$_{(9,14)}$,\emph{v}=0 & 22$_{(9,13)}$,\emph{v}=1 & 2.499E-04 & 423 & 175.6\\
230305.630  &  23$_{(8,15)}$,\emph{v}=0 & 22$_{(8,14)}$,\emph{v}=1 & 2.610E-04 & 423 & 167.4\\
230472.528  &  21$_{(4,17)}$,\emph{v}=1 & 20$_{(4,16)}$,\emph{v}=0 & 2.836E-04 & 301 & 124.2\\
230830.319  &  24$_{(2,22)}$,\emph{v}=0 & 23$_{(2,21)}$,\emph{v}=1 & 2.822E-04 & 343 & 149.8\\
230933.676  &  23$_{(3,20)}$,\emph{v}=0 & 22$_{(3,19)}$,\emph{v}=1 & 3.166E-04 & 423 & 143.3\\
230965.547  &  23$_{(7,17)}$,\emph{v}=0 & 22$_{(7,16)}$,\emph{v}=1 & 2.715E-04 & 329 & 160.2\\
231127.401  &  23$_{(7,16)}$,\emph{v}=0 & 22$_{(7,15)}$,\emph{v}=1 & 2.722E-04 & 423 & 160.2\\
231524.033  &  23$_{(6,18)}$,\emph{v}=0 & 22$_{(6,17)}$,\emph{v}=1 & 2.714E-04 & 329 & 154.1\\
231564.005  &  24$_{(1,24)}$,\emph{v}=1 & 23$_{(1,23)}$,\emph{v}=0 & 3.043E-04 & 343 & 136.8\\
231564.320  &  24$_{(0,24)}$,\emph{v}=1 & 23$_{(0,23)}$,\emph{v}=0 & 3.043E-04 & 441 & 136.8\\
232350.059  &  22$_{(10,13)}$,\emph{v}=1 & 21$_{(10,12)}$,\emph{v}=0 & 2.420E-04 & 315 & 173.8\\
232350.068  &  22$_{(10,12)}$,\emph{v}=1 & 21$_{(10,11)}$,\emph{v}=0 & 2.420E-04 & 405 & 173.8\\
232597.215  &  22$_{(9,14)}$,\emph{v}=1 & 21$_{(9,13)}$,\emph{v}=0 & 2.549E-04 & 315 & 164.6\\
232597.490  &  22$_{(9,13)}$,\emph{v}=1 & 21$_{(9,12)}$,\emph{v}=0 & 2.548E-04 & 405 & 164.6\\
232881.533  &  23$_{(6,17)}$,\emph{v}=0 & 22$_{(6,16)}$,\emph{v}=1 & 2.607E-04 & 423 & 154.3\\
232987.353  &  22$_{(8,14)}$,\emph{v}=1 & 21$_{(8,13)}$,\emph{v}=0 & 2.669E-04 & 405 & 156.3\\
233536.696  &  22$_{(5,18)}$,\emph{v}=1 & 21$_{(5,17)}$,\emph{v}=0 & 2.930E-04 & 315 & 137.7\\
233561.785  &  22$_{(7,16)}$,\emph{v}=1 & 21$_{(7,15)}$,\emph{v}=0 & 2.785E-04 & 315 & 149.1\\
233664.319  &  22$_{(7,15)}$,\emph{v}=1 & 21$_{(7,14)}$,\emph{v}=0 & 2.788E-04 & 405 & 149.1\\
234264.446  &  22$_{(6,17)}$,\emph{v}=1 & 21$_{(6,16)}$,\emph{v}=0 & 2.839E-04 & 315 & 143.0\\
235304.050  &  22$_{(6,16)}$,\emph{v}=1 & 21$_{(6,15)}$,\emph{v}=0 & 2.897E-04 & 405 & 143.1\\
235600.179  &  23$_{(2,21)}$,\emph{v}=1 & 22$_{(2,20)}$,\emph{v}=0 & 3.276E-04 & 329 & 138.7\\
235620.372  &  24$_{(4,21)}$,\emph{v}=0 & 23$_{(4,20)}$,\emph{v}=1 & 2.881E-04 & 441 & 155.4\\
235834.240  &  26$_{(1,26)}$,\emph{v}=0 & 25$_{(1,25)}$,\emph{v}=1 & 3.222E-04 & 477 & 159.3\\
235834.327  &  26$_{(0,26)}$,\emph{v}=0 & 25$_{(0,25)}$,\emph{v}=1 & 3.221E-04 & 371 & 159.3\\
244879.919  &  23$_{(6,18)}$,\emph{v}=1 & 22$_{(6,17)}$,\emph{v}=0 & 3.028E-04 & 423 & 154.4\\
245022.738  &  27$_{(1,27)}$,\emph{v}=0 & 26$_{(1,26)}$,\emph{v}=1 & 3.616E-04 & 385 & 171.4\\
245022.787  &  27$_{(0,27)}$,\emph{v}=0 & 26$_{(0,26)}$,\emph{v}=1 & 3.617E-04 & 495 & 171.4\\
246387.881  &  23$_{(6,17)}$,\emph{v}=1 & 22$_{(6,16)}$,\emph{v}=0 & 3.280E-04 & 329 & 154.6\\
250300.410  &  25$_{(10,16)}$,\emph{v}=0 & 24$_{(10,15)}$,\emph{v}=1 & 3.208E-04 & 357 & 209.0\\
250300.508  &  25$_{(10,15)}$,\emph{v}=0 & 24$_{(10,14)}$,\emph{v}=1 & 3.209E-04 & 459 & 209.0\\
250487.421  &  23$_{(5,18)}$,\emph{v}=1 & 22$_{(5,17)}$,\emph{v}=0 & 3.606E-04 & 329 & 150.3\\
250731.885  &  25$_{(9,17)}$,\emph{v}=0 & 24$_{(9,16)}$,\emph{v}=1 & 3.341E-04 & 357 & 199.8\\
250734.147  &  25$_{(9,16)}$,\emph{v}=0 & 24$_{(9,15)}$,\emph{v}=1 & 3.341E-04 & 459 & 199.8\\
251382.563  &  25$_{(8,17)}$,\emph{v}=0 & 24$_{(8,16)}$,\emph{v}=1 & 3.471E-04 & 459 & 191.6\\
251473.045  &  25$_{(6,20)}$,\emph{v}=0 & 24$_{(6,19)}$,\emph{v}=1 & 1.607E-04 & 357 & 178.4\\
251574.351  &  27$_{(2,26)}$,\emph{v}=0 & 26$_{(2,25)}$,\emph{v}=1 & 3.873E-04 & 385 & 179.3\\
251577.144  &  27$_{(1,26)}$,\emph{v}=0 & 26$_{(1,25)}$,\emph{v}=1 & 3.870E-04 & 495 & 179.3\\
\end{longtable}
\tablefoot{\\
\tablefoottext{a}{Tunnelling is observed between two equivalent equilibrium configurations and splits each rotational level into two distinct states designated as $\emph{v}=0$ and $\emph{v}=1$ (BD15).}}
}

\begin{table}[h!]
\caption{Spectroscopic data of the gGg' ethylene glycol lines displayed in Fig. \ref{fg4}}       
\label{table:B3}      
\centering          
\begin{tabular}{c c c c c c }     % 6 columns 
\hline\hline       
 frequency&\multicolumn{2}{c}{Quantum number\tablefootmark{a} }&A$_{u,l}$&g$_{up}$&E$_{up}$\\
(MHz)&J$_{K_a,K_c}$,\emph{v}(up)&J$_{K_a,K_c}$,\emph{v}(low)&(s$^{-1}$)&&(K)\\
 \hline                    
%                                     Two column figure (place early!)
217539.718  &  22$_{(2,20)}$,\emph{v}=0 & 21$_{(2,19)}$,\emph{v}=1 & 1.495E-04 & 315 & 126.6\\
218712.336  &  22$_{(3,20)}$,\emph{v}=1 & 21$_{(3,19)}$,\emph{v}=0 & 1.512E-04 & 315 & 126.7\\
220249.787  &  22$_{(2,20)}$,\emph{v}=1 & 21$_{(2,19)}$,\emph{v}=0 & 1.550E-04 & 405 & 126.7\\
229906.833  &  24$_{(2,23)}$,\emph{v}=1 & 23$_{(1,22)}$,\emph{v}=1 & 1.013E-04 & 343 & 142.7\\
231212.070  &  24$_{(1,23)}$,\emph{v}=1 & 23$_{(1,22)}$,\emph{v}=0 & 1.210E-04 & 441 & 142.7\\
231366.043  &  25$_{(1,25)}$,\emph{v}=0 & 24$_{(1,24)}$,\emph{v}=1 & 1.185E-04 & 357 & 147.0\\
231366.176  &  25$_{(0,25)}$,\emph{v}=0 & 24$_{(0,24)}$,\emph{v}=1 & 1.184E-04 & 459 & 147.0\\
232706.108  &  25$_{(0,25)}$,\emph{v}=1 & 24$_{(1,24)}$,\emph{v}=1 & 1.199E-04 & 357 & 147.1\\
232706.561  &  25$_{(1,25)}$,\emph{v}=1 & 24$_{(0,24)}$,\emph{v}=1 & 1.201E-04 & 459 & 147.1\\
233690.540  &  22$_{(4,18)}$,\emph{v}=1 & 21$_{(4,17)}$,\emph{v}=0 & 1.118E-04 & 405 & 134.2\\
234047.938  &  25$_{(1,25)}$,\emph{v}=1 & 24$_{(1,24)}$,\emph{v}=0 & 1.225E-04 & 459 & 147.1\\
234048.079  &  25$_{(0,25)}$,\emph{v}=1 & 24$_{(0,24)}$,\emph{v}=0 & 1.226E-04 & 357 & 147.1\\
234150.007  &  23$_{(8,15)}$,\emph{v}=0 & 22$_{(8,14)}$,\emph{v}=1 & 1.039E-04 & 423 & 166.1\\
236771.321  &  23$_{(6,17)}$,\emph{v}=0 & 22$_{(6,16)}$,\emph{v}=1 & 1.139E-04 & 423 & 153.1\\
236845.214  &  23$_{(7,17)}$,\emph{v}=1 & 22$_{(7,16)}$,\emph{v}=0 & 1.102E-04 & 423 & 159.0\\
236864.590  &  23$_{(3,20)}$,\emph{v}=1 & 22$_{(3,19)}$,\emph{v}=0 & 1.058E-04 & 329 & 142.4\\
245205.575  &  25$_{(2,23)}$,\emph{v}=0 & 24$_{(3,22)}$,\emph{v}=0 & 1.714E-04 & 459 & 160.6\\
245215.148  &  24$_{(7,18)}$,\emph{v}=0 & 23$_{(7,17)}$,\emph{v}=1 & 1.232E-04 & 441 & 170.8\\
245238.282  &  24$_{(12,12)}$,\emph{v}=1 & 23$_{(12,11)}$,\emph{v}=0 & 1.016E-04 & 441 & 216.4\\
245238.282  &  24$_{(12,13)}$,\emph{v}=1 & 23$_{(12,12)}$,\emph{v}=0 & 1.016E-04 & 343 & 216.4\\
245424.852  &  24$_{(11,14)}$,\emph{v}=1 & 23$_{(11,13)}$,\emph{v}=0 & 1.073E-04 & 343 & 205.3\\
245424.853  &  24$_{(11,13)}$,\emph{v}=1 & 23$_{(11,12)}$,\emph{v}=0 & 1.073E-04 & 441 & 205.3\\
245681.514  &  24$_{(10,15)}$,\emph{v}=1 & 23$_{(10,14)}$,\emph{v}=0 & 1.126E-04 & 343 & 195.2\\
245681.559  &  24$_{(10,14)}$,\emph{v}=1 & 23$_{(10,13)}$,\emph{v}=0 & 1.126E-04 & 441 & 195.2\\
245742.900  &  24$_{(6,19)}$,\emph{v}=0 & 23$_{(6,18)}$,\emph{v}=1 & 1.280E-04 & 441 & 164.8\\
246042.955  &  24$_{(9,16)}$,\emph{v}=1 & 23$_{(9,15)}$,\emph{v}=0 & 1.176E-04 & 343 & 186.0\\
246044.081  &  24$_{(9,15)}$,\emph{v}=1 & 23$_{(9,14)}$,\emph{v}=0 & 1.176E-04 & 441 & 186.0\\
250230.085  &  19$_{(4,15)}$,\emph{v}=0 & 18$_{(3,15)}$,\emph{v}=1 & 1.043E-04 & 351 & 102.1\\
250473.822  &  25$_{(4,22)}$,\emph{v}=0 & 24$_{(4,21)}$,\emph{v}=1 & 1.059E-04 & 357 & 166.5\\
\hline                  
\end{tabular}
\tablefoot{\\
\tablefoottext{a}{Tunnelling is observed between two equivalent equilibrium configurations and splits each rotational level into two distinct states designated $\emph{v}=0$ and $\emph{v}=1$ (BD15).}}
\end{table}

\clearpage

%------------------------------------------------------------------
% ---Contamination---
%-----------------------------------------------------------------
\section{Contamination}
\label{sec:app3}

\begin{figure}[h!]
\includegraphics[width=\hsize]{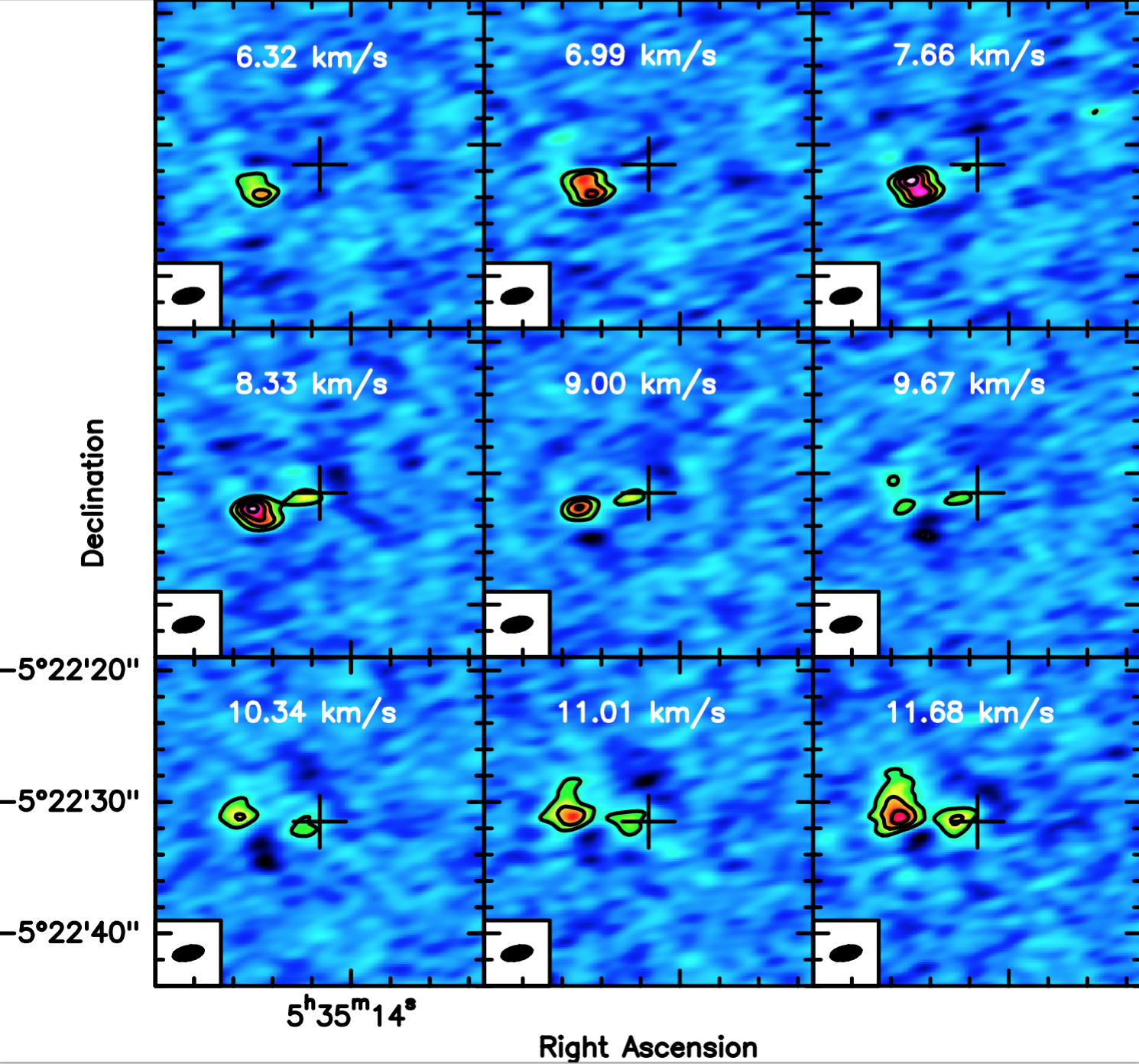}
\includegraphics[width=\hsize]{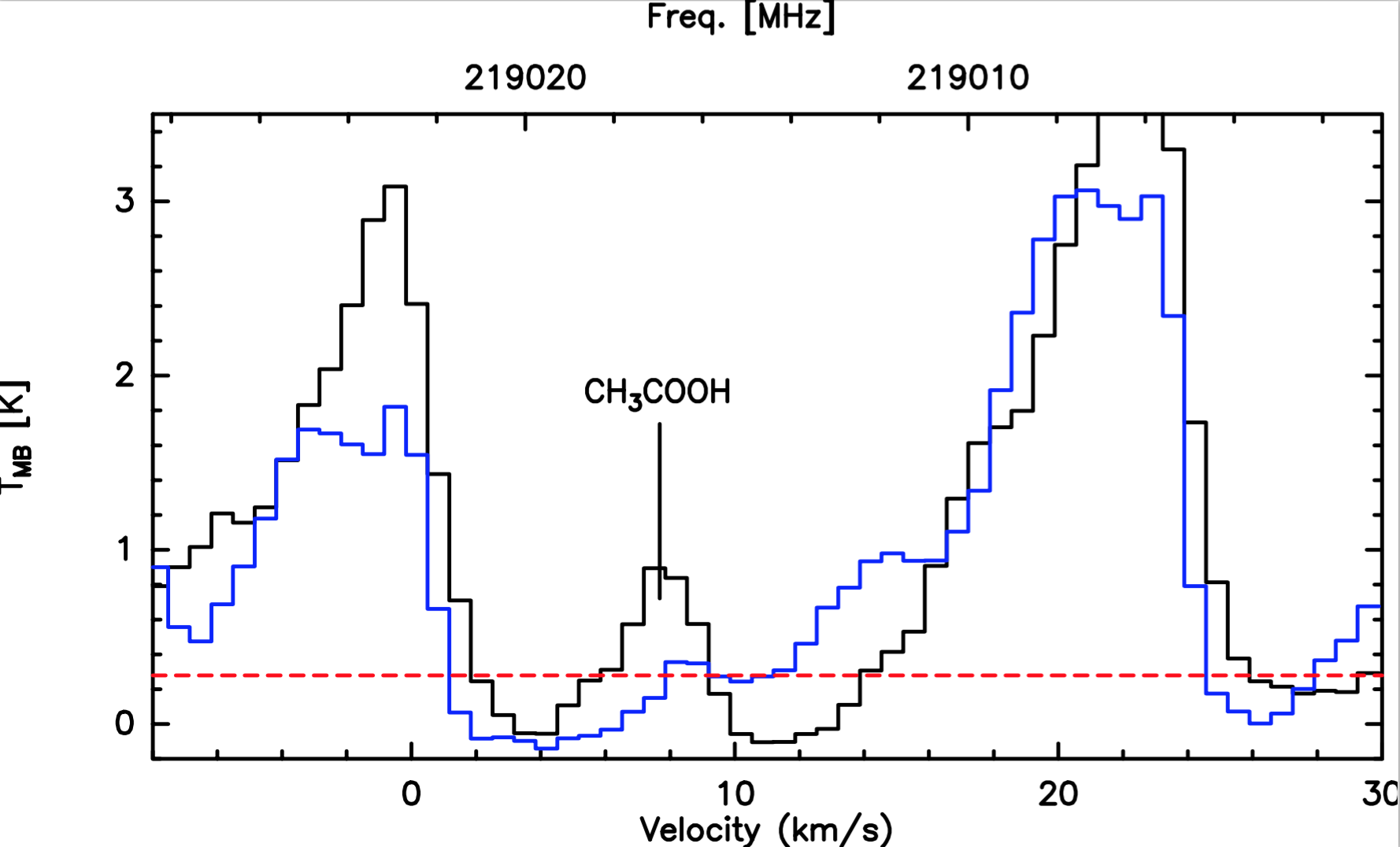}
\caption{{\it Top Panel:} CH$_3$COOH channel emission maps at 219016~MHz. {\it Bottom Panel}: Spectra centred at 219016~MHz. The black spectrum is taken in direction of the EGP emission peak while the blue one is taken in direction of the northwest clump which contaminates the CH$_3$COOH emission maps displayed here as well as in Figure~\ref{fg1}. The red dashed line shows the 3$\sigma$ noise level of the spectrum taken in direction of the northwest clump. 
}
\label{fgC}
\end{figure}

Figure~\ref{fgC} shows that the acetic acid emission at 219016~MHz is partially contaminated by the emission from an unidentified species towards the northwest region from the EGP peak.

%---------------

%------------------------------------------------------------------
% ---Comparison with HCOOCH3 ---
%-----------------------------------------------------------------
\section{Comparison with the HCOOCH$_3$ emission }
\label{sec:app4}

\begin{figure}[h!]
\includegraphics[angle=0,width=7.5cm]{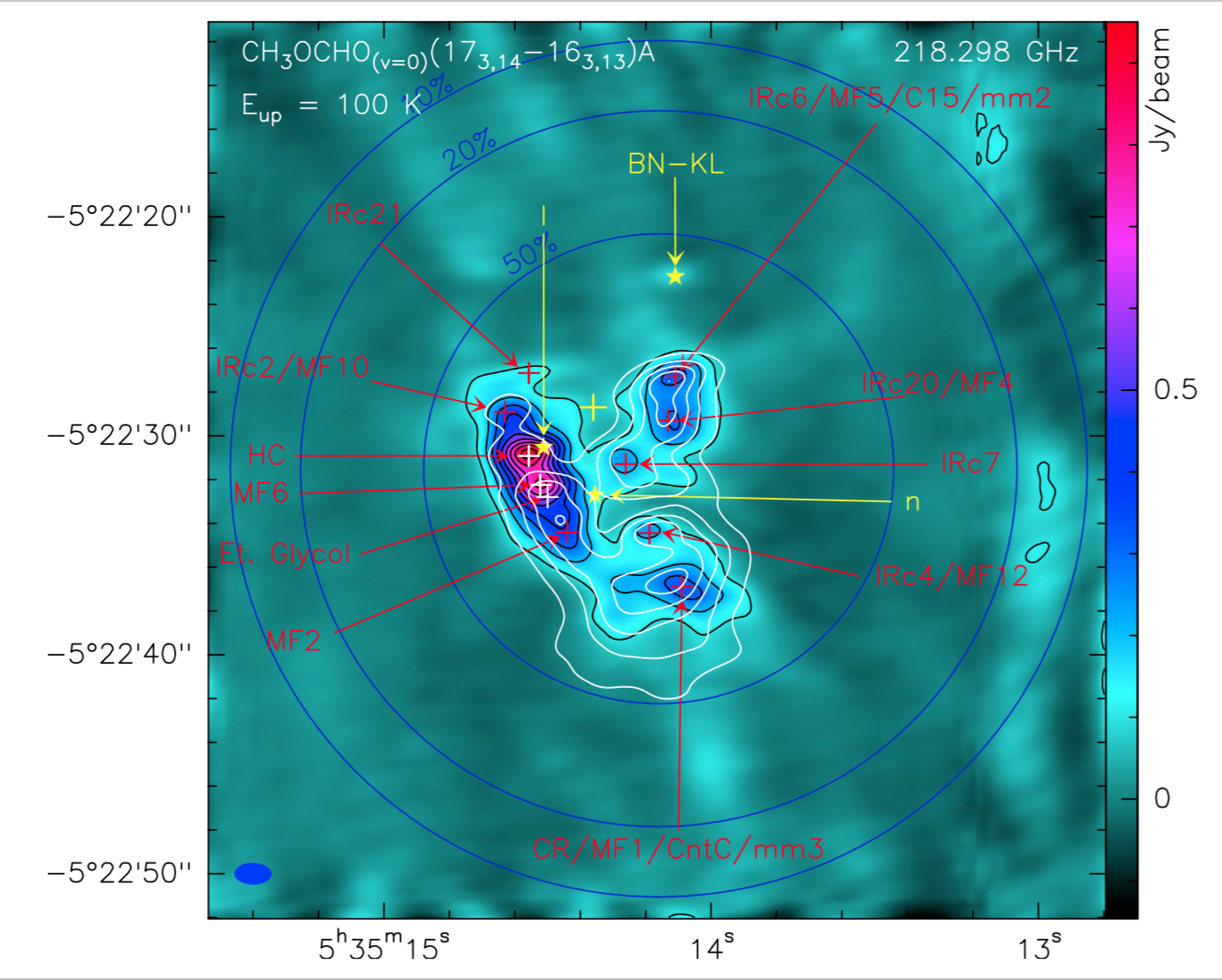}
\caption{Continuum emission at 1.3mm (color) overlaid with the HCOOCH$_3$ (write contours) emission at 218298~MHz. Positions of the sources analysed in our Paper I are also given.}
\label{fgD}
\end{figure}

Figure~\ref{fgD} illustrates the fact the distribution of the emission associated with the acetic acid and the ethylene glycol molecules differs from that of typical O-bearing species, such as methyl formate (HCOOCH$_3$) within Orion-KL.

%---------------

%-----------------------------------------------------------------
% --RING AND EMISSION ---
%-----------------------------------------------------------------
\section{HC$_3$N molecular ring and acetic acid and ethylene glycol emission }
\label{sec:app5}
\begin{figure}[h!]
\includegraphics[width=7cm]{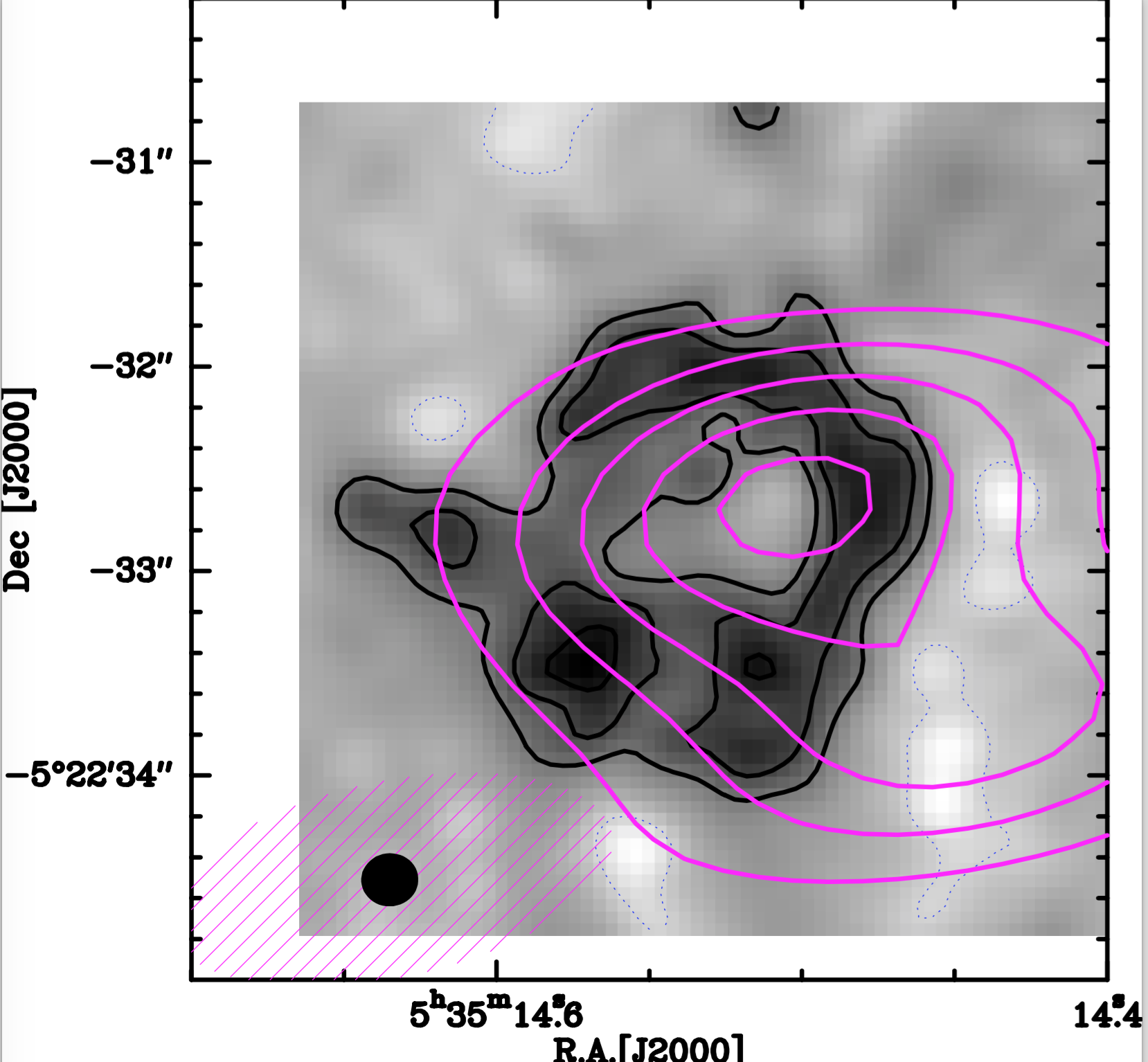}
\includegraphics[width=7cm]{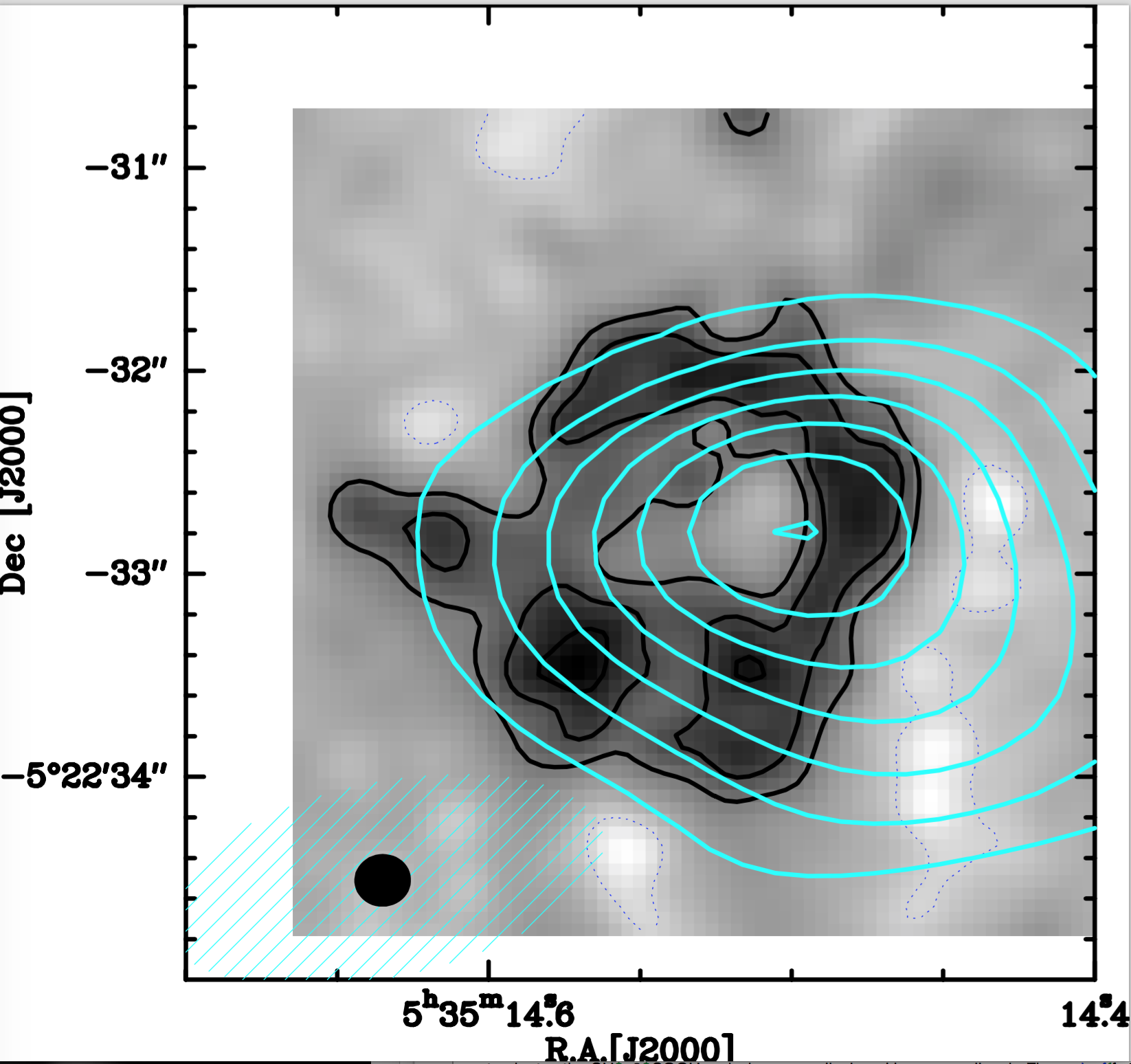}
\includegraphics[width=7cm]{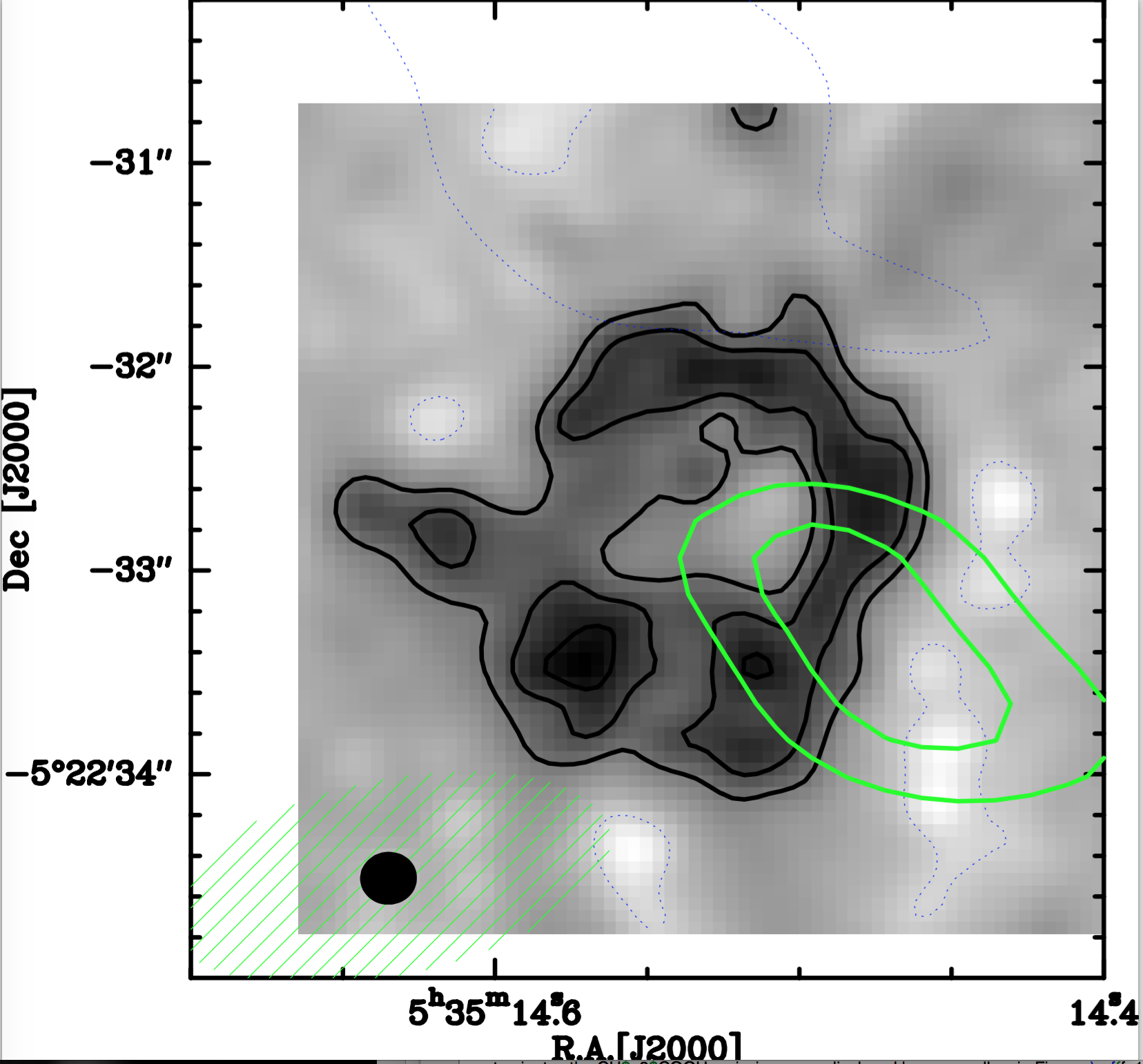}
\caption{ALMA observations of the HC$_3$N emission at 354.69~GHz \citep[in grey scale, see][]{Wright:2017} overlaid with the emission of acetic acid (purple contours, top panel), aGg$^{\prime}$--ethylene glycol (cyan contours, middle panel) and gGg$^{\prime}$--ethylene glycol (green contours, bottom panel). 
The ALMA synthesized beams are shown as the black circles for the HC$_3$N data \citep[][]{Wright:2017} and as colored ellipses for our data.}
\label{fgE}
\end{figure}

The three panels of Figure~\ref{fgE} show the HC$_3$N ring-like structure emission \citep{Wright:2017} overlaid with the emission of acetic acid, aGg$^{\prime}$--ethylene glycol and gGg$^{\prime}$--ethylene glycol  towards the Orion Kleinmann--Low nebula.

%===============================================================

\end{appendix}
%
%===============================================================
%
%===============================================================
\end{document}